\documentclass[12pt]{article}

\newcommand{\beq}{\begin{equation}}
\newcommand{\eeq}{\end{equation}}
\newcommand{\beqa}{\begin{eqnarray}}
\newcommand{\eeqa}{\end{eqnarray}}
\newcommand{\beqar}{\begin{eqnarray*}}
\newcommand{\eeqar}{\end{eqnarray*}}

\newcommand{\al}{\alpha}
\newcommand{\be}{\beta}

\def\non          {\nonumber}
\def\ha           {\mbox{$\frac{1}{2}$}}

\def\Tr           {\mbox{\rm Tr}\,}
\def\STr          {\mbox{\rm STr}\,}

\def\cd           {{\cdot}}
\def\ran          {\rangle}
\def\lan          {\langle}
\def\fsH    {H\!\!\!\!/\,}
\def\fsC    {C\!\!\!\!/\,}

\newcommand{\eps}{\epsilon}
\newcommand{\ga}{\gamma}

\newcommand{\inn}{\!\cdot\!}

\newcommand{\lam}{\lambda}

\newcommand{\z}{\zeta}

\newcommand{\labell}[1]{\label{#1}} 
\newcommand{\reef}[1]{(\ref{#1})}
\newcommand\prt{\partial}
\newcommand\veps{\varepsilon}

\newcommand\cL{{\cal L}}
\newcommand\cD{{\cal D}}

\newcommand\bz{\bar{z}}

\begin{document}
\baselineskip 17pt%
\begin{titlepage}
\vspace*{.8mm}%
  \halign{#\hfil         \cr
         CERN-PH-TH/2012-342\cr
           } 
\vspace*{8mm}%


\center{ {\bf \Large  Shedding light on new Wess-Zumino couplings with their corrections to
all orders in  alpha-prime
}}\vspace*{1mm} \centerline{{\Large {\bf  }}}
\vspace*{3mm}
\begin{center}
{Ehsan Hatefi }$\footnote{E-mail:ehatefi@ictp.it}$

\vspace*{0.6cm}{ {\it
International Centre for Theoretical Physics\\
 Strada Costiera 11, Trieste, Italy
  \\
  and
  
 Theory Group, Physics Department, CERN\\
 CH-1211, Geneva 23, Switzerland
}}

\vspace*{.2cm}
\end{center}
\begin{center}{\bf Abstract}\end{center}
\begin{quote}
Motivated by arXiv:1203.5553, we continue to match super string amplitudes with their own effective field theory. We carry out within full details the computations of the complete form of the amplitude of one closed string Ramond-Ramond field and three SYM vertex operators, namely one gauge field and two scalar fields in type IIB(A) super string theories. Making use of the recent two gauge and two scalar couplings to all orders of $\alpha'$ , we produce  all the infinite gauge poles for $p=n$ case (with $n$ as RR field strength's rank and $p$ as the dimension of a  D$_p$-brane). Proceeding with direct and full S-Matrix calculations, we are able to produce even all
the infinite gauge poles in u-channel for $p-2=n$ case in field theory as well. New couplings for $p-2=n$ case  with their all order $\alpha'$ corrections are discovered. In addition, we explain  how to find out all the infinite scalar poles of this amplitude in $s,t$-channels and produce them for $p+2=n$ case.
By comparing all of the contact terms of this amplitude, we obtain several new couplings  with their  higher derivative corrections for $p+2=n,p=n$ cases. These new couplings are neither inside Myers'terms nor within pull-back/Taylor expansions. Finally we comment on some related issues.

\end{quote}
\end{titlepage}


\section{Introduction}

D$_p$-branes have been centering in String theory on both theoretical and phenomenological approaches for a while.
 For diverse values of $p$ (where $p$ is spatial dimension of a D$_p$-brane) and also in both type IIA and IIB string theories, they have been known as the sources of closed string Ramond-Ramond field  \cite{Polchinski:1995mt},\cite{Witten:1995im,Polchinski:1996na}.

\vskip 0.1in

By computing some of the couplings of D$_p$-branes to closed string modes, some great information have been obtained. We address various examples such as the Ads/CFT correspondence, gauge theory and black holes.
Concerning Ramond-Ramond couplings \cite{Li:1995pq,Green:1996dd}, diverse phenomena such as \cite{Witten:1995gx,Douglas:1995bn}, realizing K-theory in terms of D-branes \cite{Minasian:1997mm,Witten:1998cd} and Myers effect \cite{Myers:1999ps,Taylor:1999gq,Taylor:1999pr} have been discovered.

\vskip 0.1in

In order to review string duality  \cite{pol 12} is highly proposed. To observe information on the world volume of a D$_p$-brane and in particular to deal with both Dirac-Born-Infeld and Chern-Simons effective actions we refer to \cite{Hatefi:2010ik,Tseytlin:1999dj,Tseytlin:1997csa} and all references
therein. In order to encounter the effective action only for a bosonic D$_p$-brane, \cite{Leigh:1989jq} should be highlighted. It is widely understood that for multiple D$_p$-branes, references \cite{Myers:1999ps,Taylor:1999pr} are the main ones to look for bosonic action.

\vskip 0.1in

    In order to see super symmetric action, one might search about some special references in \cite{Howe:2006rv} and \cite{Cederwall:1996pv}.

\vskip 0.1in

Basically one has to emphasize the fact that the higher derivative corrections of stable and unstable branes are not involved in those effective actions, namely the only way for obtaining the closed form of all corrections is indeed scattering computations. To have all corrections, recent attempts in detail have been carried out. Morever, to discover the higher derivative  corrections for stable branes, namely  four field strengths' corrections to all orders in $\alpha'$ \cite{Hatefi:2010ik} must be taken into account, also there  we have shown that to $\alpha'^4$ order computations are indeed consistent with literature \cite{Bilal:2001hb,Chandia:2003sh,Barreiro:2005hv}. To achieve two gauge field and two scalar fields ' corrections  again to all orders of $\alpha'$, \cite{Hatefi:2012ve} is suggested. Finally a pattern and a universal prescription for
all BPS branes, including  corrections to four covariant derivative of scalar fields has been found in \cite{Hatefi:2012rx}.

\vskip 0.1in

 Although in this paper we are interested in finding new couplings of gauge/scalars in the
 background of one closed string Ramond-Ramond field to all orders of $\alpha'$, arguing effective actions for stable branes, addressing some recent works for unstable branes, such as  \cite{Hatefi:2012wj} and \cite{Garousi:2008ge} is highly recommended. In particular the effective action of brane-anti brane to all orders of $\alpha'$ for two gauge and two tachyons have been derived in \cite{Garousi:2007fk}.  Recently remarks on the effective action of brane anti brane with all their $\alpha'$ corrections in  \cite{Hatefi:2012cp} have been made. Note that we do not review Wess-Zumino effective action here,
 but in order to follow all needed couplings, section 5 of \cite{Hatefi:2012wj} and references
 \cite{Li:1995pq}, \cite{Douglas:1995bn} and \cite{Green:1996dd}
 might be studied to pursue them for different values of $p$ and $n$, where $n$ is the rank of the field strength of the closed string Ramond-Ramond field.

\vskip 0.3in

 Given several goals, involving dualities of Ads/CFT \cite{Hatefi:2012bp} and some exact relation between open and closed strings inside the Ads/CFT, examining new method for higher point string amplitudes is indeed necessary.

 \vskip 0.2in

Two extremely important facts which must be really highlighted are :

\vskip 0.1in

 1) In order to find all new couplings/contact interactions  to all orders of $\alpha'$ one has to have the complete form of the amplitudes. Note that the result of the amplitudes at leading orders is not very useful as we comment it in detail in this paper.

 \vskip 0.1in

 2)Once we are dealing with open-closed amplitudes T-duality transformation is not very effective and in fact direct computations of those amplitudes are inevitable. Even we want to work out tree level amplitude, the appearance of closed string RR makes the calculations so complicated. It is definitely realized that in loop  computations applying T-duality is really subtle \cite{Park:2009ki}. For example in \cite{Hatefi:2012rx} we have shown that
 it is not possible to derive $<V_C V_\phi V_\phi V_\phi> $ from $<V_C V_A V_A V_\phi> $. In particular we have seen that the terms including momentum of closed string RR in transverse directions $p^i,p^j$ are not appeared in $<V_C V_A V_A V_\phi> $.

 \vskip 0.1in

Now we address some of the motivations for the long computations of this paper. The first is to realize the closed form of new Wess-Zumino couplings to all orders in $\alpha'$. It was argued in \cite{Hatefi:2012wj} that  for the amplitudes including scalar fields and closed string Ramond-Ramond field, field theory vertices must be obtained just by three different methods, basically either through Myers' terms \cite{Myers:1999ps} or pull-back approach or Taylor expansion. However in this paper within detail we will show that there are some new couplings which do come from none of them. These new couplings for sure do not come from Myers' terms or Taylor expansion so this is a very strong evidence in favor of modifying pull-back (see also \cite{Hatefi:2012ve,Hatefi:2012rx}).

 \vskip 0.1in

The other motivation is to get more data to see whether or not essentially we can write down the general form of DBI and Wess-Zumino effective actions. To check up to some orders \cite{Koerber:2002zb,Keurentjes:2004tu,Denef} may be useful to look at.

 \vskip 0.1in

The third reason for including $<V_C V_A V_\phi V_\phi> $ is indeed its direct relation to dielectric effect.
As an example $N^3$ entropy of M5 branes was expressed in terms of dielectric effect in \cite{Hatefi:2012ab}, for the other applications and for various configurations in M-theory see \cite{Hatefi:2012ac,McOrist:2012yc}.
Definitely the results of this paper on new couplings will provide fundamental steps for future outcomes such as all orders dielectric effect and various research topics on world volume dynamics of branes \cite{Hatefi:2012bp}.

 \vskip 0.1in

Two remarkable issues have been addressed. The first one is related to taking integrals for five point open-closed strings \cite{Fotopoulos:2001pt} in which for the first time has been applied in \cite{Garousi:2007fk} and we were able to find several new couplings in the world volume of brane- anti brane systems. The second important fact is related to Wick-like formula \cite{Liu:2001qa}. Making use of this formula  and by generalizing that in \cite{Hatefi:2010ik}, we can now simply derive the correlation function between two spin operators in the presence of several currents and fermion fields. To deal with the integrations on some higher point functions, we suggest \cite{Stieberger:2009hq}.

 \vskip 0.1in

 Dielectric effect does have various applications such as resolving some singularities in Ads/CFT by making use of closed string RR field to actually  polarize $D_{3}$-brane (more details can be seen in \cite{Polchinski:2000uf,Pilch:2000fu,Aharony:2000cw}). The importance of this effect inside M-theory  is argued in \cite{Bena:2000zb,Bena:2000fz}, even it is more discussed in stabilizing the sources of RR in some particular backgrounds \cite{Bachas:2000ik} relating to Wess-Zumino-Witten model, $Ads_m\times S^n$, fuzzy gravitons in some space-time \cite{Li:2000bd,Grisaru:2000zn} or in terms of  their gravity duals \cite{Trivedi:2000mq}.

 \vskip 0.1in

 Given these applications, we are going to explore all new couplings between one closed string Ramond-Ramond field $(C)$ and some SYM vertex operators, namely three open strings (basically two scalars and one gauge field) in the IIB(A) super string theories. This $<V_C V_A V_\phi V_\phi> $ was introduced in
 \cite{Garousi:2000ea} and in that paper making use of hyper geometric function the authors just were able to find the first simple pole of the amplitude, however, the complete form of the amplitude is unknown which we are going to find it out. Apart from that the authors  in \cite{Garousi:2000ea} have produced just the first scalar t-channel and the first gauge $(s+t+u)$-channel pole.

 \vskip 0.1in

  Using new techniques for five point amplitudes ( see Appendix B of  \cite{Hatefi:2012wj}) and making use of Wick-like formula we find the complete form of the amplitude of $<V_C V_A V_\phi V_\phi> $.
In addition to obtaining several new couplings to all orders of $\alpha'$ which we will address in detail, we are going to show that the amplitude not only has gauge u-channel pole but also it has infinite u-channel poles which have been overlooked in \cite{Garousi:2000ea} and could not be derived there because of not having the complete form of the amplitude.

Also, making use of the all order two gauge two scalar couplings  that appeared in the very recent paper \cite{Hatefi:2012ve},
 we will find out the infinite gauge $(t+s+u)$-channel poles and infinite scalar $(t,s)$-channel poles as well.

 Then we go on further and discover several new couplings for $p+2=n$ case. In particular by applying direct S-matrix computations and discovering the complete and closed form of the correlators of $<V_C V_A V_\phi V_\phi> $    we derive  all new interactions to all orders in $\alpha'$ for  $p-2=n,p=n$ cases as well.

 \section { Complete form of the  $CA\phi\phi$ amplitude to all orders of $\alpha'$ }

Here we are going to explore the S-matrix elements of another physically 4-point and technically 5-point function. Namely we do want to investigate in detail the amplitude of 3 BPS branes (2 scalar and one gauge fields) and one closed string RR field. Motivation for computing such a long computation is indeed checking all infinite couplings two gauge and two scalar fields which are recently discovered \cite{Hatefi:2012ve} and also trying to figure out how one can reproduce all infinite massless gauge and scalar fields for different values of $p$ and $n$.
  To compute a S-matrix element, one should clarify  the picture of the vertex operators in an appropriate way. It would be nice to refer to some new works on BPS branes \cite{Hatefi:2012ve},\cite{Hatefi:2012rx} and \cite{Barreiro:2012aw}.

 \vskip 0.2in

 Note that since we are looking for $<V_C V_A V_\phi V_\phi> $ some parts of the calculations  are shared with
 \cite{Hatefi:2010ik,Hatefi:2012ve,Garousi:2000ea} but definitely all of the contact terms at all orders are new results and different from the previous results. It is important to highlight the following point.
 Due to $C$-field, it is not possible to obtain all contact terms of this paper from
 $<V_C V_A V_A V_{\phi}> $ by applying T-duality transformation, because we will observe here that all the terms involving $p^i,p^j$  are not present in $<V_C V_A V_A V_{\phi}> $.

 \vskip 0.2in

We must use the vertex operators according to the fact that for the disk level amplitudes (which we are dealing with) total charge has to be -2 .


Taking into account super ghost charges, we may write down the amplitude of one gauge, two open scalar fields and one closed string RR in terms of some kinds of special correlators

\begin{eqnarray}
{\cal A}^{CA\phi\phi} & \sim & \int dx_{1}dx_{2}dx_{3}dzd\bar{z}\,
  \lan V_{A}^{(0)}{(x_{1})}
V_{\phi}^{(0)}{(x_{2})}V_\phi^{(0)}{(x_{3})}
V_{RR}^{(-\frac{3}{2},-\frac{1}{2})}(z,\bar{z})\ran,\labell{sstring}\eeqa

Since we are performing disk level amplitudes , all open strings must be put on the boundary of disk rather, RR has to be replaced in the middle of disk .

 Depending on the various picture of the strings , One should pick up the following vertex operators.
The vertex operators  are known as \footnote{In
string calculations, we used to set $\alpha'=2$.} \beqa
V_{\phi}^{(0)}(x) &=& \xi_{i}\bigg(\partial
X^i(x)+\alpha'iq\cd\psi\psi^i(x)\bigg)e^{\alpha'iq.X(x)},
\nonumber\\
V_{A}^{(0)}(x) &=& \xi_{a}\bigg(\partial
X^a(x)+\alpha'ik\cd\psi\psi^a(x)\bigg)e^{\alpha'ik.X(x)},
\nonumber\\
V_{A}^{(-2)}(x) &=& e^{-2\phi(x)}V_{A}^{(0)}(x),
\nonumber\\
V_{A}^{(-1)}(y) &=&\xi_a\psi^a(y) e^{-\phi(y)} e^{\alpha'ik\cd X(y)},
\nonumber\\
V_{RR}^{(-\frac{1}{2},-\frac{1}{2})}(z,\bar{z})&=&(P_{-}\fsH_{(n)}M_p)^{\al\be}e^{-\phi(z)/2}
S_{\al}(z)e^{i\frac{\alpha'}{2} p\cd X(z)}e^{-\phi(\bar{z})/2} S_{\be}(\bar{z})
e^{ i\frac{\alpha'}{2} p\cd D \cd X(\bar{z})},
\nonumber\\
V_{RR}^{(-\frac{3}{2},-\frac{1}{2})}(z,\bar{z})&=&(P_{-}\fsC_{(n)} M_p)^{\al\be}e^{-3\phi(z)/2}
S_{\al}(z)e^{i\frac{\alpha'}{2} p\cd X(z)}e^{-\phi(\bar{z})/2} S_{\be}(\bar{z})
e^{ i\frac{\alpha'}{2} p\cd D \cd X(\bar{z})},
\label{d4Vs}
\eeqa
 $q,k$ are scalar field and gauge field 's momenta
which do  satisfy the following conditions
 $k^2=q^2=0$ and $k_i.\xi_j=0$.
 The definitions of projector and field strength of RR are
 \begin{displaymath}
 \fsH_{(n)} = \frac{a
 _n}{n!}H_{\mu_{1}\ldots\mu_{n}}\ga^{\mu_{1}}\ldots
 \ga^{\mu_{n}}\nonumber\\
\ ,P_{-} = \ha (1-\ga^{11})
\non\end{displaymath}
  $n$ is odd/even number for type IIB/IIA theory. To see more notation \cite{Hatefi:2010ik} is recommended.
 The simplest way to do this computation is as follows
 \begin{eqnarray}
{\cal A}^{CA\phi\phi} & \sim & \int dx_{1}dx_{2}dx_{3}dzd\bar{z}\,
  \lan V_{A}^{(-1)}{(x_{1})}
V_{\phi}^{(0)}{(x_{2})}V_\phi^{(0)}{(x_{3})}
V_{RR}^{(-\frac{1}{2},-\frac{1}{2})}(z,\bar{z})\ran,\labell{25sstring}\eeqa
Making use of the standard  correlators for $X^{\mu},\psi^{\mu}, \phi$  as follows
\begin{eqnarray}
\lan X^{\mu}(z)X^{\nu}(w)\ran & = & -\frac{\alpha'}{2}\eta^{\mu\nu}\log(z-w) , \non \\
\lan \psi^{\mu}(z)\psi^{\nu}(w) \ran & = & -\frac{\alpha'}{2}\eta^{\mu\nu}(z-w)^{-1} \ ,\non \\
\lan\phi(z)\phi(w)\ran & = & -\log(z-w) \ .
\labell{prop}\end{eqnarray}
also introducing $x_{4}\equiv\ z=x+iy$ , $x_{5}\equiv\bz=x-iy$, the final form of the amplitude with just taking $\Tr(\lam_1\lam_2\lam_3)$ ordering and with the closed form of the
correlators reaches to
\beqa {\cal A}^{CA\phi\phi}&\sim& \int
 dx_{1}dx_{2}dx_{3}dx_{4} dx_{5}\,
(P_{-}\fsH_{(n)}M_p)^{\al\be}\xi_{1a}\xi_{2i}\xi_{3j}x_{45}^{-1/4}(x_{14}x_{15})^{-1/2}\nonumber\\&&
\times(I_1+I_2+I_3+I_4)\Tr(\lam_1\lam_2\lam_3),\labell{125}\eeqa where
$x_{ij}=x_i-x_j$. Having taken Wick theorem,  one gets the correlators as
\beqa
I_1&=&{<:e^{\alpha'ik_1.X(x_1)}:\partial X^i(x_2)e^{\alpha' ik_2.X(x_2)}
:\partial X^j(x_3)e^{\alpha' ik_3.X(x_3)}:e^{i\frac{\alpha'}{2} p\cd X(x_4)}:e^{ i\frac{\alpha'}{2} p\cd D \cd X(x_5)} :>}
 \  \non \\&&\times{<:S_{\al}(x_4):S_{\be}(x_5):\psi^a(x_1):>},\nonumber\\
I_2&=&{<:e^{\alpha' ik_1.X(x_1)}:e^{\alpha' ik_2.X(x_2)}
:\partial X^j(x_3)e^{\alpha' ik_3.X(x_3)}:e^{i\frac{\alpha'}{2} p\cd X(x_4)}:e^{ i\frac{\alpha'}{2} p\cd D \cd X(x_5)}:>}
 \  \non \\&&\times{<:S_{\al}(x_4):S_{\be}(x_5)::\psi^a(x_1):\alpha' ik_2.\psi\psi^{i}(x_2)>},\nonumber\\
 I_3&=&{<: e^{\alpha' ik_1.X(x_1)}:\partial X^i(x_2)e^{\alpha' ik_2.X(x_2)}
:e^{\alpha' ik_3.X(x_3)}:e^{i\frac{\alpha'}{2} p\cd X(x_4)}:e^{ i\frac{\alpha'}{2} p\cd D \cd X(x_5)}:>}
 \  \non \\&&\times{<:S_{\al}(x_4):S_{\be}(x_5)::\psi^a(x_1):\alpha' ik_3.\psi\psi^{j}(x_3)>},\nonumber\\
 I_4&=&{<: e^{\alpha' ik_1.X(x_1)}:e^{\alpha' ik_2.X(x_2)}
:e^{\alpha' ik_3.X(x_3)}:e^{i\frac{\alpha'}{2} p\cd X(x_4)}:e^{ i\frac{\alpha'}{2} p\cd D \cd X(x_5)}:>}
 \  \non \\&&\times{<:S_{\al}(x_4):S_{\be}(x_5):\psi^a(x_1)
:\alpha' ik_{2}\cd\psi\psi^i(x_2):\alpha' ik_{3}\cd\psi\psi^j(x_3):>}.
\label{i1234}
\eeqa
Needless to remind that the following correlation has been achieved by working out the generalized form of Wick-like \cite{Hatefi:2010ik}
\beqa
I_5^{iba}&=&<:S_{\al}(x_4):S_{\be}(x_5):\psi^a(x_1):\psi^b\psi^i(x_2):>\nonumber\\
&=&\bigg\{(\Gamma^{iba}C^{-1})_{\alpha\beta}
+\frac{\alpha' Re[x_{14}x_{25}]}{x_{12}x_{45}}\bigg(-\eta^{ab}(\gamma^{i}C^{-1})_{\alpha\beta}\bigg)\bigg\}
\nonumber\\&&\times2^{-3/2}x_{45}^{1/4}(x_{24}x_{25})^{-1}(x_{14}x_{15})^{-1/2}.
\label{68}\eeqa
Having used the arguments mentioned in \cite{Hatefi:2010ik}, the following correlator can be easily gained:
\beqa
I_6^{jciba}&=&<:S_{\al}(x_4):S_{\be}(x_5)::\psi^a(x_1):\psi^b\psi^i(x_2):\psi^c\psi^j(x_3)>\nonumber\\
&=&\bigg\{(\Gamma^{jciba}C^{-1})_{{\alpha\beta}}+\alpha' r_1\frac{Re[x_{14}x_{25}]}{x_{12}x_{45}}+\alpha' r_2\frac{Re[x_{14}x_{35}]}{x_{13}x_{45}}+\alpha' r_3\frac{Re[x_{24}x_{35}]}{x_{23}x_{45}}+(\alpha'^2) r_4\nonumber\\&&\times\bigg(\frac{Re[x_{24}x_{35}]}{x_{23}x_{45}}\bigg)^{2}
+(\alpha'^2) r_5\bigg(\frac{Re[x_{14}x_{25}]}{x_{12}x_{45}}\times\frac{Re[x_{24}x_{35}]}{x_{23}x_{45}}\bigg)+(\alpha'^2) r_6\bigg(\frac{Re[x_{14}x_{35}]}{x_{13}x_{45}}\nonumber\\&&\times\frac{Re[x_{24}x_{35}]}{x_{23}x_{45}}\bigg)
\bigg\}2^{-5/2}x_{45}^{5/4}(x_{24}x_{25}x_{34}x_{35})^{-1}(x_{14}x_{15})^{-1/2},\label{hhiop}\eeqa
where
\beqa
r_1&=&\bigg(-\eta^{ab}(\Gamma^{jci}C^{-1})_{\alpha\beta}\bigg),\nonumber\\
r_2&=&\bigg(-\eta^{ac}(\Gamma^{jib}C^{-1})_{\alpha\beta}
\bigg),\nonumber\\
r_3&=&\bigg(\eta^{bc}(\Gamma^{jia}C^{-1})_{\alpha\beta}
+\eta^{ij}(\Gamma^{cba}C^{-1})_{\alpha\beta}\bigg),\nonumber\\
r_4&=&\bigg((-\eta^{bc}\eta^{ij})(\gamma^{a}C^{-1})_{\alpha\beta}\bigg),\nonumber\\
r_5&=&\bigg((-\eta^{ab}\eta^{ij})(\gamma^{c}C^{-1})_{\alpha\beta}\bigg),\nonumber\\
r_6&=&\bigg((\eta^{ac}\eta^{ij})(\gamma^{b}C^{-1})_{\alpha\beta} \bigg).
\eeqa
Having regarded all those correlators in our amplitude, we find the closed form of these S-matrix elements as :
\beqa
{\cal A}^{CA\phi\phi}&\!\!\!\!\sim\!\!\!\!\!&\int dx_{1}dx_{2} dx_{3}dx_{4}dx_{5}(P_{-}\fsH_{(n)}M_p)^{\al\be}I\xi_{1a}\xi_{2i}\xi_{3j}x_{45}^{-1/4}(x_{14}x_{15})^{-1/2}\nonumber\\&&\times\bigg(I_7^a(-\eta^{ij}x_{23}^{-2}+a^j_1a^i_2)+a^j_1a^{ia}_3+a^i_2a^{ja}_4-\alpha'^2 k_{2b}k_{3c}I_6^{jciba}\bigg)\Tr(\lam_1\lam_2\lam_3)\labell{amp3},\eeqa
where  $I_6^{kbjai}$ is given in \reef{hhiop} and
\beqa
I&=&|x_{12}|^{\alpha'^2 k_1.k_2}|x_{13}|^{\alpha'^2 k_1.k_3}|x_{14}x_{15}|^{\frac{\alpha'^2}{2} k_1.p}|x_{23}|^{\alpha'^2 k_2.k_3}|x_{24}x_{25}|^{\frac{\alpha'^2}{2} k_2.p}
|x_{34}x_{35}|^{\frac{\alpha'^2}{2} k_3.p}|x_{45}|^{\frac{\alpha'^2}{4}p.D.p},\nonumber\\
a^j_1&=&ip^{j}\frac{x_{54}}{x_{34}x_{35}},\nonumber\\
a^i_2&=&ip^{i}\frac{x_{54}}{x_{24}x_{25}},\nonumber\\
a^{ia}_3&=&\alpha' ik_{2b}I_5^{iba},\nonumber\\
a^{ja}_4&=&\alpha' ik_{3c}2^{-3/2}x_{45}^{1/4}(x_{34}x_{35})^{-1}(x_{14}x_{15})^{-1/2} \nonumber\\&&\times\bigg\{(\Gamma^{jca}C^{-1})_{\alpha\beta}
+\frac{\alpha' Re[x_{14}x_{35}]}{x_{13}x_{45}}\bigg(-\eta^{ac}(\gamma^{j}C^{-1})_{\alpha\beta}\bigg)\bigg\}
,\nonumber\\
I_7^a&=&<:S_{\al}(x_4):S_{\be}(x_5):\psi^a(x_1):>=2^{-1/2}x_{45}^{-3/4}(x_{14}x_{15})^{-1/2}
(\gamma^{a}C^{-1})_{\alpha\beta}.
\label{isas}
\eeqa
Now we are ready to show that the amplitude is written such that $SL(2,R)$ transformation holds.

We apply a special gauge fixing which is different from the ones that appeared in \cite{Garousi:2000ea}, that is why we can find out the general form of the amplitude. Basically we just fixed the positions of all three massless open strings in $(0,1,\infty)$ and carry out all integrations by making use of the integrals obtained in \cite{Hatefi:2012wj}.
 The other fact which has been widely used is indeed introducing  the following Mandelstam variables
 \beqa
s&=&-\frac{\alpha'}{2}(k_1+k_3)^2,\qquad t=-\frac{\alpha'}{2}(k_1+k_2)^2,\qquad u=-\frac{\alpha'}{2}(k_2+k_3)^2.
\nonumber\eeqa

 Eventually we try to find out the final form of the amplitude \reef{amp3} as
\beqa {\cal A}^{CA\phi\phi}&=&{\cal A}_{1}+{\cal A}_{2}+{\cal A}_{3}+{\cal A}_{4}+{\cal A}_{5}+{\cal A}_{6}
+{\cal A}_{7}+{\cal A}_{8}+{\cal A}_{9}+{\cal A}_{10}\labell{11u}\eeqa
where
\beqa
{\cal A}_{1}&\!\!\!\sim\!\!\!&-2^{-1/2}\xi_{1a}\xi_{2i}\xi_{3j}
\bigg[k_{3c}k_{2b}\Tr(P_{-}\fsH_{(n)}M_p\Gamma^{jciba})-k_{2b}p^j\Tr(P_{-}\fsH_{(n)}M_p\Gamma^{iba})\nonumber\\&&-k_{3c}p^i\Tr(P_{-}\fsH_{(n)}M_p\Gamma^{jca})+p^ip^j\Tr(P_{-}\fsH_{(n)}M_p\gamma^{a})\bigg]
L_1,
\nonumber\\
{\cal A}_{2}&\sim&2^{-1/2}
\bigg\{-
2\xi_{1}.k_{2}k_{3c}\xi_{3j}\xi_{2i}\Tr(P_{-}\fsH_{(n)}M_p \Gamma^{jci})\bigg\}L_2\nonumber\\
{\cal A}_{3}&\sim&2^{-1/2}
\bigg\{\xi_{1a}\xi_{2i}\xi_{3j}\Tr(P_{-}\fsH_{(n)}M_p \Gamma^{jia})\bigg\}L_{22}\nonumber\\
{\cal A}_{4}&\sim&2^{-1/2}
\bigg\{
2k_{3}.\xi_{1}k_{2b}\xi_{3j}\xi_{2i}\Tr(P_{-}\fsH_{(n)}M_p \Gamma^{jib})\bigg\}L_3\nonumber\\
\nonumber\\
{\cal A}_{5}&\sim&2^{-1/2}
\bigg\{2\xi_{3}.\xi_{2}k_{2b}k_{3c}\xi_{1a}\Tr(P_{-}\fsH_{(n)}M_p \Gamma^{cba})\bigg\}L_5\nonumber\\
{\cal A}_{6}&\sim& 2^{1/2}L_{2}\bigg\{p^j\xi_1.k_2\xi_{2i}\xi_{3j}\Tr(P_{-}\fsH_{(n)}M_p\gamma^i)
\bigg\}
\nonumber\\
{\cal A}_{7}&\sim&-2^{-1/2}L_3\bigg\{
2k_{3}.\xi_1p^i\xi_{3j}\xi_{2i}\Tr(P_{-}\fsH_{(n)}M_p\gamma^j)\bigg\}
\nonumber\\
{\cal A}_{8}&\sim&2^{1/2}L_6\bigg\{2k_2.\xi_1 k_{3c}\Tr(P_{-}\fsH_{(n)}M_p\gamma^c)
(-s\xi_2.\xi_3)\bigg\}.
\nonumber\\
{\cal A}_{9}&\sim&2^{1/2}L_6\bigg\{2k_3.\xi_{1}k_{2b}\Tr(P_{-}\fsH_{(n)}M_p\gamma^b)
(-t\xi_2.\xi_3)\bigg\}
\nonumber\\
{\cal A}_{10}&\sim&2^{1/2}L_6\bigg\{\xi_{1a}\Tr(P_{-}\fsH_{(n)}M_p\gamma^a)
(ts\xi_3.\xi_2)\bigg\}
\labell{480}\eeqa
where the functions
 $L_1,L_2,L_{22},L_3,L_5,L_6$ are appeared in the following
\beqa
L_1&=&(2)^{-2(t+s+u)+1}\pi{\frac{\Gamma(-u+\frac{1}{2})
\Gamma(-s+\frac{1}{2})\Gamma(-t+\frac{1}{2})\Gamma(-t-s-u+1)}
{\Gamma(-u-t+1)\Gamma(-t-s+1)\Gamma(-s-u+1)}},\nonumber\\
L_2&=&(2)^{-2(t+s+u)}\pi{\frac{\Gamma(-u+1)
\Gamma(-s+1)\Gamma(-t)\Gamma(-t-s-u+\frac{1}{2})}
{\Gamma(-u-t+1)\Gamma(-t-s+1)\Gamma(-s-u+1)}}
\nonumber\\
L_{22}&=&(2)^{-2(t+s+u)}\pi{\frac{\Gamma(-u+1)
\Gamma(-s+1)\Gamma(-t+1)\Gamma(-t-s-u+\frac{1}{2})}
{\Gamma(-u-t+1)\Gamma(-t-s+1)\Gamma(-s-u+1)}}
\nonumber\\
L_3&=&(2)^{-2(t+s+u)}\pi{\frac{\Gamma(-u+1)
\Gamma(-s)\Gamma(-t+1)\Gamma(-t-s-u+\frac{1}{2})}{\Gamma(-u-t+1)
\Gamma(-t-s+1)\Gamma(-s-u+1)}}
,\nonumber\\
L_5&=&(2)^{-2(t+s+u)}\pi{\frac{\Gamma(-u)
\Gamma(-s+1)\Gamma(-t+1)\Gamma(-t-s-u+\frac{1}{2})}
{\Gamma(-u-t+1)\Gamma(-t-s+1)\Gamma(-s-u+1)}}
,\nonumber\\
L_6&=&(2)^{-2(t+s+u)-1}\pi{\frac{\Gamma(-u+\frac{1}{2})
\Gamma(-s+\frac{1}{2})\Gamma(-t+\frac{1}{2})\Gamma(-t-s-u)}
{\Gamma(-u-t+1)\Gamma(-t-s+1)\Gamma(-s-u+1)}},
\label{Ls}
\eeqa

We could actually simplify the final result more as follows :
\beqa {\cal A}^{C A\phi\phi}&=&{\cal A}_{1}+{\cal A}_{2}+{\cal A}_{3}+{\cal A}_{4},\labell{141u}\eeqa
where
\beqa
{\cal A}_{1}&\!\!\!\sim\!\!\!&2^{-1/2}4\xi_{1a}\xi_{2i}\xi_{3j}(t+s+u)L_1'
\bigg[k_{3c}k_{2b}\Tr(P_{-}\fsH_{(n)}M_p\Gamma^{jciba})-k_{2b}p^j\Tr(P_{-}\fsH_{(n)}M_p\Gamma^{iba})\nonumber\\&&-k_{3c}p^i\Tr(P_{-}\fsH_{(n)}M_p\Gamma^{jca})+p^ip^j\Tr(P_{-}\fsH_{(n)}M_p\gamma^{a})\bigg]
\nonumber\\
{\cal A}_{2}&\sim&2^{-1/2}L_2'
\bigg\{-2us\xi_{1}.k_{2}\xi_{2i}k_{3c}\xi_{3j}\Tr(P_{-}\fsH_{(n)}M_p \Gamma^{jci})-ust\xi_{1a}\xi_{2i}\xi_{3j}\Tr(P_{-}\fsH_{(n)}M_p \Gamma^{jia})
\nonumber\\&&
+2ut k_{3}.\xi_{1}k_{2b}\xi_{3j}\xi_{2i}\Tr(P_{-}\fsH_{(n)}M_p \Gamma^{jib})
+2us \xi_{2i}p^j\xi_1.k_2\xi_{3j}\Tr(P_{-}\fsH_{(n)}M_p\gamma^i)\nonumber\\&&
-2ut \xi_{3j}p^i\xi_1.k_3\xi_{2i}\Tr(P_{-}\fsH_{(n)}M_p\gamma^j)\bigg\}\nonumber\\
{\cal A}_{3}&\sim&2^{1/2}\Tr(P_{-}\fsH_{(n)}M_p\gamma^a)\xi_3.\xi_2 L_1'\bigg[ts\xi_{1a}
-2tk_3.\xi_{1}k_{2a}
-2sk_2.\xi_{1}k_{3a}\bigg]
\nonumber\\
{\cal A}_{4}&\sim&2^{-1/2}L_2'
\bigg\{ 2st\xi_{3}.\xi_{2}k_{2b}k_{3c}\xi_{1a}\Tr(P_{-}\fsH_{(n)}M_p \Gamma^{cba})\bigg\}.
\labell{47988}\eeqa

\vskip 0.1in

where the functions
 $L_1',L_2'$ now defined as
 \vskip 0.1in

\beqa
L_1'&=&(2)^{-2(t+s+u)-1}\pi{\frac{\Gamma(-u+\frac{1}{2})
\Gamma(-s+\frac{1}{2})\Gamma(-t+\frac{1}{2})\Gamma(-t-s-u)}
{\Gamma(-u-t+1)\Gamma(-t-s+1)\Gamma(-s-u+1)}},\nonumber\\
L_2'&=&(2)^{-2(t+s+u)}\pi{\frac{\Gamma(-u)
\Gamma(-s)\Gamma(-t)\Gamma(-t-s-u+\frac{1}{2})}{\Gamma(-u-t+1)
\Gamma(-t-s+1)\Gamma(-s-u+1)}}
\label{L1L2}
\eeqa

\vskip 0.2in

One important test of our amplitude is indeed applying Ward identity. Replacing $\xi_{1a}\rightarrow k_{1a}$, one believes that  all parts of  amplitude become zero. Notice that our amplitude  makes sense for $n=p-2,  n=p+2$ and $p=n$ cases.

Given the facts that we are dealing with 5-point super string computations for all massless strings and we are applying momentum conservation just for longitudinal direction,  it is expected to get the same relation as appeared in \cite{Hatefi:2010ik,Hatefi:2012ve}. Thus the following relation holds

\beqa
s+t+u=-p_ap^a.
\labell{cons}\eeqa

As argued in \cite{Hatefi:2012rx}, the expansion must be done by sending all three Mandelstam variables to zero.

It is worth taking the fact that both $L_1',L_2'$ are symmetrized
 in terms of $(u,t,s)$ and this provides some confusions to indeed derive the general form of our expansions, however the field theory is a very useful guide in order for obtaining desired expansions. Since we are carrying out technically 5-point function for all massless particles, the expansions and the coefficients are the same as appeared in the amplitude of one Ramond-Ramond and three massless scalar fields\cite{Hatefi:2012rx}, however, the  terms that appeared in  those amplitudes are really different from the terms of $<V_C V_A V_{\phi} V_{\phi} >$ and this is one the reasons for performing explicit computations. The corrected expansions are
\beqa
 L_1' &=&-{\frac{\pi^{5/2}}{2}}\left( \sum_{n=0}^{\infty}c_n(s+t+u)^n\right.
\left.+\frac{\sum_{n,m=0}^{\infty}c_{n,m}[s^n t^m +s^m t^n]}{(t+s+u)}\right.\nonumber\\
&&\left.+\sum_{p,n,m=0}^{\infty}f_{p,n,m}(s+t+u)^p[(s+t)^{n}(st)^{m}]\right),\label{expansion11}
\eeqa

\beqa
su L_2' &=&-\pi^{3/2}\bigg(\sum_{n=-1}^{\infty}b_n\bigg(\frac{1}{t}(u+s)^{n+1}\bigg)
            +\sum_{p,n,m=0}^{\infty}e_{p,n,m}t^{p}(su)^{n}(s+u)^m\bigg)
 \label {expansion226}
\eeqa
Finally in order to obtain the suitable expansions for $st L_2'$ and $tu L_2'$ , one must replace $t \leftrightarrow u$ in \reef{expansion226} and $t \leftrightarrow s$ accordingly such that
\beqa
tu L_2' &=&-\pi^{3/2}\bigg(\sum_{n=-1}^{\infty}b_n\bigg(\frac{1}{s}(u+t)^{n+1}\bigg)
            +\sum_{p,n,m=0}^{\infty}e_{p,n,m}s^{p}(tu)^{n}(t+u)^m\bigg)
 \label {expansion227}
\eeqa
\beqa
ts L_2' &=&-\pi^{3/2}\bigg(\sum_{n=-1}^{\infty}b_n\bigg(\frac{1}{u}(s+t)^{n+1}\bigg)
            +\sum_{p,n,m=0}^{\infty}e_{p,n,m}u^{p}(ts)^{n}(t+s)^m\bigg)
 \label {expansion228}
\eeqa

In order to produce all massless poles for different values of $p$ and $n$ one has to know some of the coefficients in those expansions :
\beqa
&&b_{-1}=1,\,b_0=0,\,b_1=\frac{1}{6}\pi^2,\,b_2=2\z(3),c_0=0,c_1=\frac{\pi^2}{6},e_{0,0,1}=\frac{1}{3}\pi^2\labell{hash},\\
&&e_{2,0,0}=e_{0,1,0}=2\z(3),e_{1,0,0}=\frac{1}{6}\pi^2,e_{1,0,2}=\frac{19}{60}\pi^4,e_{1,0,1}=e_{0,0,2}=6\z(3),\nonumber\\
&&c_2=-2\z(3),
\,c_{1,1}=\frac{\pi^2}{6},\,c_{0,0}=\frac{1}{2},c_{3,1}=c_{1,3}=\frac{2}{15}\pi^4,c_{2,2}=\frac{1}{5}\pi^4,f_{0,1,0}=-\frac{1}{3}\pi^2\nonumber\\
&&c_{1,0}=c_{0,1}=0,
c_{3,0}=c_{0,3}=0\,
,\,c_{2,0}=c_{0,2}=\frac{\pi^2}{6},c_{1,2}=c_{2,1}=-4\z(3),c_{4,0}=c_{0,4}=\frac{1}{15}\pi^4 ,\nonumber
\eeqa
The important point here is that, ${ L}_1'$ for our amplitude   $<V_CV_A V_{\phi} V_{\phi}>$  must have infinite massless gauge but not scalar poles in the $(t+s+u)$-channel and this is unlike $<V_CV_A V_A V_{\phi}>$, which had infinite massless scalar poles. The other point which must be mentioned before carrying out field theory computations is that,
${ L}_2' $ must have either infinite massless  scalar poles in  $t$,$s$ channels or infinite massless gauge poles in $u$-channels which we take care of them in a closed form in the next sections.

\subsection{Infinite massless gauge  poles for $p-2=n$ case }

In this section, we are going to explore all infinite u-channel gauge poles with new Wess-Zumino couplings. These poles have been overlooked in \cite{Garousi:2000ea}.
By applying $st L_2'$ expansion into  ${\cal A}_{4}$ amplitude in \reef{141u} one should have got  all massless gauge poles in string amplitude as
\beqa
-\pi^2 \mu_p (2\xi_{3}.\xi_{2}k_{2b}k_{3c}\xi_{1a}) \frac{16}{(p-2)!}\eps^{a_{0}\cdots a_{p-3}cba}H_{a_{0}\cdots a_{p-3}}
\sum_{n=-1}^{\infty}b_n\bigg(\frac{1}{u}(t+s)^{n+1}\bigg)\Tr(\lam_1\lam_2\lam_3)
\label{upoles}\eeqa
where trace has been taken and we just kept all poles in the expansion of $stL_2'$. Note that we normalized the amplitude by multiplying a coefficient of $2^{1/2}\pi^{1/2}\mu_p$.

The important point should be highlighted is that, \reef{upoles} has been anti symmetrized in terms of both scalars and this leads to the conclusion that amplitude must be non-vanished just for non Abelian gauge group.

One test of this part of amplitude is, taking into account Ward identity for the gauge field. Thus by replacing $\xi_{1a} \rightarrow k_{1a}$, making use of momentum conservation and applying physical state condition for RR ($p^a \eps^{a_{0}\cdots a_{p-3}cba}=0$), we observe that \reef{upoles} does vanish.

 \vskip 0.2in

The related Feynman rule in field theory side  for $p-2=n$ case is

\beqa
{\cal A}&=&V^a_{\alpha}(C_{p-3},A_1,A)G^{ab}_{\alpha\beta}(A)V^b_{\beta}(A,\phi_2,\phi_3),\labell{amp644}
\eeqa


 The needed vertex $V^a_{\alpha}(C_{p-3},A_1,A)$ in field theory must be obtained by taking this Chern-Simons coupling

\beqa
S_1&=& i\lambda^2\mu_p\int d^{p+1}\sigma \quad \Tr(C_{(p-3)}\wedge F \wedge F)
\label{s2}
\eeqa

Note that in the above action $F_{ab}=\partial^a A^b-\partial^b A^a-i[A^a,A^b]$ and  $\lambda= 2\pi\alpha'$, however all commutators must be neglected as we are looking for the coupling between one RR-$(p-3)$ form and two gaue fields. Integration by parts are also taken  such that

\beqa
V^a_{\alpha}(C_{p-3},A_1,A)&=&\lambda^2\mu_p {1\over (p-2)!} \eps^{a_0\cdots a_{p-1}a}H_{a_0\cdots a_{p-3}}\xi_{1a_{p-2}}k_{a_{p-1}}
\nonumber\eeqa

 As it becomes clear from \reef{upoles} the amplitude has infinite poles. The vertex of $V^b_{\beta}(A,\phi_2,\phi_3)$ should be derived from the kinetic term of scalar fields in DBI action $[\frac{\lambda^2}{2}\Tr(D^a\phi^iD_a\phi_i)]$ as follows

\beqa
V_{\beta}^{b}(A,\phi_2,\phi_3)&=&i\lambda^2T_p
  \xi_2.\xi_3 (k_2-k_3)^b \Tr(\lambda_2\lambda_3\lambda_\beta)
\nonumber\\
G_{\alpha\beta}^{ab}(\phi)&=&\frac{-i}{\lambda^2T_p}\frac{\delta^{ab}
\delta_{\alpha\beta}}{k^2}\,\,\, ,
\label{ver137}
\eeqa
Consider $k^2=-(k_2+k_3)^2=u$  in the above propagator.

Kinetic term of scalar field  indeed has been fixed so definitely there is no correction to all kinetic terms such as kinetic term of  scalars. Also notice that massless poles here are simple massless poles thus neither do they  get corrected. In addition to that, by considering
\reef{s2} we could produce just the first simple gauge pole out of infinite poles.

 \vskip 0.1in

 Having got this remarkable fact, we come to the point that, in order to produce all infinite gauge poles one has to find out all related corrections to $i\lambda^2\mu_p\int d^{p+1}\sigma  C_{(p-3)}\wedge F \wedge F$ as

 \beqa
 S_2&=& i\lambda^2\mu_p\int d^{p+1}\sigma   \quad \sum_{n=-1}^{\infty}b_n (\alpha')^{n+1}\quad C_{(p-3)}\wedge D_{a_0\cdots a_n} F \wedge D^{a_0\cdots a_n} F
\eeqa
 for more explanations see \cite{Hatefi:2010ik}. By setting these corrections, we are indeed able to produce all massless gauge poles to all orders of $\alpha'$. Let us write down the corrected form of the needed vertex to all orders as

 \beqa
V^a_{\alpha}(C_{p-3},A_1,A)&=&\frac{\lambda^2\mu_p}{(p-2)!}(\eps)^{a_0\cdots a_{p-1}a}(H^{(p-2)})_{a_0\cdots a_{p-3}}\xi_{1a_{p-2}}k_{a_{p-1}}\nonumber\\&&\times\Tr(\lam_1\lambda_\alpha)\sum_{n=-1}^{\infty}b_n(\alpha'k_1.k)^{n+1}
\label{ver12}\eeqa

 Having replaced \reef{ver12},\reef{ver137} into \reef{amp644} we may write down the result as
\beqa
{\cal A}&=&\mu_p(2\pi\alpha')^{2}\frac{1}{(p-2)!u}\Tr(\lam_1\lam_2\lam_3)\eps^{a_{0}\cdots a_{p-1}a}H_{a_{0}\cdots a_{p-3}}\xi_{1a_{p-2}}\sum_{n=-1}^{\infty}b_n\bigg(\frac{\alpha'}{2}\bigg)^{n+1}(s+t)^{n+1}
\nonumber\\&&\times\bigg[ 2k_{2a} k_{3a_{p-1}}\xi_2.\xi_3\bigg]\label{99}\eeqa
 Thus \reef{99} can exactly produce all ifinite massless gauge poles which we were looking for in \reef{upoles}. Indeed we have precisely produced all u-channel poles in this section. This is the other new result of this paper.

\section{New Couplings for BPS-branes for $n=p-2$ case}

In this section, by comparing direct result of string amplitude , we are going to discover new Wess-Zumino couplings for $n=p-2$ case
at leading order and generalize them to actually construct all their higher order corrections as well.

\vskip .1in

To start, we rewrite the explicit form of string amplitude for this case as
\beqa
A_4&=&2\pi^{1/2}\mu_p st\xi_{3}.\xi_{2}k_{2b}k_{3c}\xi_{1a}\Tr(P_{-}\fsH_{(n)}M_p \Gamma^{cba})L_2'
\nonumber\eeqa
Extracting the trace and applying $st L_2'$ expansion we get
\beqa
A_4&=& -2\xi_{3}.\xi_{2}k_{2b}k_{3c}\xi_{1a}\pi^2\mu_p\frac{16}{(p-2)!}\eps^{a_{0}\cdots a_{p-3}cba}H_{a_{0}\cdots a_{p-3}} \nonumber\\&&\times
\bigg(\sum_{n=-1}^{\infty}b_n\bigg(\frac{1}{u}(t+s)^{n+1}\bigg)
            +\sum_{p,n,m=0}^{\infty}e_{p,n,m}u^{p}(st)^{n}(s+t)^m\bigg)\label {newcop}\eeqa
In the last section,  and in particular in \reef{99}, comparing with string theory amplitude, we have produced all infinite u-channel poles in field theory. Now in order for gaining new couplings with exact coefficients, what we have to take into account is indeed the second term in \reef{newcop} ,that is,
\beqa
A_4^{CA\phi\phi}&=& -2\xi_{3}.\xi_{2}k_{2b}k_{3c}\xi_{1a}\pi^2\mu_p\frac{16}{(p-2)!}\eps^{a_{0}\cdots a_{p-3}cba}H_{a_{0}\cdots a_{p-3}} \nonumber\\&&\times
\bigg(\sum_{p,n,m=0}^{\infty}e_{p,n,m}u^{p}(st)^{n}(s+t)^m\bigg)\label {newcop3}\eeqa


Note that \reef{newcop3} by itself does satisfy Ward identity , namely if we replace $\xi_{1a}$ to $k_{1a}$ and apply the momentum conservation
along the world volume and in particular consider the physical state condition for the RR $p^a\eps^{a_{0}\cdots a_{p-1}a}=0$, we come to the fact that
this part of the amplitude should be written just in terms of a new Wess-Zumino coupling which must have the following structure

\beqa
\int_{\sum_{p+1}}d^{p+1}\sigma  \quad  \Tr(C_{p-3}\wedge F \wedge D\phi^i \wedge D\phi_i)\label{newnew}\eeqa

It is written such that, it covers  the world volume space and also satisfies anti-symmetrization with respect to the interchange of scalar field's momenta. Let us  apply $e_{1,0,0}=\frac{\pi^2}{6}$ and
$e_{0,0,1}=\frac{\pi^2}{3}$
to \reef{newnew} and produce the first non-zero couplings as $S_3$ and $S_4$ then generalize all orders in $\alpha'$ higher derivative corrections :

\beqa
S_{3}&=&\frac{\lambda^3\mu_p\pi}{12}\int d^{p+1}\sigma {1\over (p-3)!}(\veps^v)^{a_0\cdots a_{p}}
\left(\frac{\alpha'}{2}\right)
\nonumber\\&&\times C^{(p-3)}_{a_0\cdots a_{p-4}}\Tr\bigg( F_{a_{p-3}a_{p-2}} (D^aD_a)  \bigg[D_{a_{p-1}}\phi^i D_{a_{p}}\phi_i\bigg]\bigg)
\labell{hderv22}
\eeqa
and

\beqa
S_{4}&=&\frac{\lambda^3\mu_p\pi}{6}\int d^{p+1}\sigma
 \left(\alpha'\right)
\Tr\bigg(C_{p-3}\wedge D^{b_1} F\wedge  D_{b_1}  \bigg[  D\phi^i \wedge  D\phi_i\bigg]\bigg)
\labell{hderv36}
\eeqa
\vskip .1in

It is not difficult to investigate that, in order to produce \reef{newcop3}, the closed form of higher derivative corrections to all orders of $\alpha'$ must be taken as follows

\beqa
S_{5}&=&\frac{\lambda^3\mu_p}{2\pi}\int d^{p+1}\sigma \sum_{p,n,m=0}^{\infty}
e_{p,n,m}\left(\alpha'\right)^{2n+m}\left(\frac{\alpha'}{2}\right)^{p}
\Tr\bigg(C_{p-3}\wedge D^{b_1}\cdots D^{b_{m}}D^{a_1}\cdots D^{a_{2n}}F\wedge \nonumber\\&& (D^aD_a)^p D_{b_1}\cdots D_{b_{m}} \bigg[ D_{a_1}\cdots D_{a_n}D\phi^i \wedge D_{a_{n+1}}\cdots D_{a_{2n}}D\phi_i\bigg]\bigg)
\labell{hderv36}
\eeqa

\section{Infinite massless gauge poles for $p=n$ case}

The goal for  this section is to show that pure super Yang-Mills (SYM) couplings (infinite two gauge and two scalar couplings in \cite{Hatefi:2012ve}) will give rise the same infinite  gauge poles in $<V_C V_A V_{\phi} V_{\phi}>$ as well. Extracting the trace and considring $L_1'$ expansion inside of the third part of the amplitude $(A_3)$, we rewrite all infinite massless gauge poles of the amplitude for  $p=n$ case  as the following:
\beqa
{\cal A}_{3}&=&\pi ^{3}\mu_p \xi_3.\xi_2 \frac{16}{ p!}
\eps^{a_{0}\cdots a_{p-1}a}H^{(p)}_{a_{0}\cdots a_{p-1}}\bigg[ts\xi_{1a}
- 2tk_3.\xi_{1}k_{2a}
-2sk_2.\xi_{1}k_{3a}\bigg]\nonumber\\&&\times
\left( \frac{\sum_{n,m=0}^{\infty}c_{n,m}[s^n t^m +s^m t^n]}{(t+s+u)}\right)
\labell{nb2}\eeqa

with $2^{1/2}\pi^{1/2}\mu_{p}$ becomes normalisation factor to match with field theory side.
The following Feynman rule
must be taken into account for this case.

\beqa
{\cal A}&=&V_{\alpha}^{a}(C_{p-1},A)G_{\alpha\beta}^{ab}(A)V_{\beta}^{b}(A,A_1,
\phi_2,\phi_3),\labell{amp5491}
\eeqa


First of all let us talk about the chern -simons coupling, namely we want to gain  $V_{\alpha}^{a}(C_{p-1},A)$ by taking the known coupling
\beqa
2\pi\alpha'\mu_p\int d^{p+1}\sigma \Tr\left(C_{p-1}\wedge F\right)\,
 \eeqa

such that
\beqa
V_{\alpha}^{a}(C_{p-1},A)&=&i(2\pi\alpha')\mu_p\frac{1}{(p)!} \eps^{a_0\cdots a_{p-1}a}
 H^{(p)}_{a_0\cdots a_{p-1}}\Tr(\lambda_{\alpha})\nonumber\\
G_{\alpha\beta}^{ab}(A) &=&\frac{i\delta_{\alpha\beta}\delta^{ab}}{T_p(2\pi\alpha')^2
k^2}=\frac{i\delta_{\alpha\beta}\delta^{ab}}{T_p(2\pi\alpha')^2
(t+s+u)}.
\labell{Fey2}
\eeqa

In order to produce all infinite gauge poles for this particular case
one has to know SYM couplings between one off-shell gauge and one on-shell gauge and two on-shell scalar fields at leading order

\beqa
&&- \frac{T_p(2\pi\alpha')^4}{2}{\rm STr}
\left(D_a\phi^iD^b\phi_iF^{ac}F_{bc}-\frac{1}{4}
(D_a\phi^i D^a\phi_iF^{bc}F_{bc})\right).\labell{a011}
\eeqa

 and in particular, we need to make use of their higher derivative corrections to all orders of $\alpha'$ which are recently discovered in \cite{Hatefi:2012ve}:

\beqa
(2\pi\alpha')^4\frac{1}{ 2 \pi^2}T_p\left(\alpha'\right)^{n+m}\sum_{m,n=0}^{\infty}(\cL_{1}^{nm}+\cL_{2}^{nm}+\cL_{3}^{nm}),\labell{highder}\eeqa
\beqa
&&\cL_{1}^{nm}=-
\Tr\left(\frac{}{}a_{n,m}\cD_{nm}[D_a \phi^i D^b \phi_i F^{ac}F_{bc}]+ b_{n,m}\cD'_{nm}[D_a \phi^i F^{ac} D^b \phi_i F_{bc}]+h.c.\frac{}{}\right),\nonumber\\
&&\cL_{2}^{nm}=-\Tr\left(\frac{}{}a_{n,m}\cD_{nm}[D_a \phi^i D^b \phi_i F_{bc}F^{ac}]+\frac{}{}b_{n,m}\cD'_{nm}[D_a \phi^i F_{bc} D^b \phi_i F^{ac}]+h.c.\frac{}{}\right),\nonumber\\
&&\cL_{3}^{nm}=\frac{1}{2}\Tr\left(\frac{}{}a_{n,m}\cD_{nm}[D_a \phi^i D^a \phi_i F^{bc}F_{bc}]+\frac{}{}b_{n,m}\cD'_{nm}[D_a \phi^i F_{bc} D^a \phi_i F^{bc}]+h.c\frac{}{}\right),\nonumber\eeqa
where the higher derivative operators
$D_{nm} $ and $ D'_{nm}$ are defined \cite{Hatefi:2010ik} as
\beqa
\cD_{nm}(EFGH)&\equiv&D_{b_1}\cdots D_{b_m}D_{a_1}\cdots D_{a_n}E  F D^{a_1}\cdots D^{a_n}GD^{b_1}\cdots D^{b_m}H,\nonumber\\
\cD'_{nm}(EFGH)&\equiv&D_{b_1}\cdots D_{b_m}D_{a_1}\cdots D_{a_n}E   D^{a_1}\cdots D^{a_n}F G D^{b_1}\cdots D^{b_m}H.\nonumber
\eeqa
The first thing to note is that, in order to obtain the vertex of one off-shell gauge and one gauge and two scalars on-shell , one should have taken into account two possible orderings  as below:
\beqa
\Tr(\lambda_2\lambda_3\lambda_1\lambda_{\beta}), \quad\quad\quad \Tr(\lambda_2\lambda_3\lambda_{\beta}\lambda_1)
\eeqa
where $\beta$ has to be Abelian. As an example if we consider $\Tr\left(a_{n,m}\cD_{nm}[D_a \phi^i D^b \phi_i F^{ac}F_{bc}]\right)$ the resulted vertex is
\beqa
a_{n,m}(k.k_2)^m(k_1.k_2)^n \xi_2.\xi_3 I_{10}\nonumber\\
+a_{n,m}(k.k_2)^n(k_1.k_2)^m \xi_2.\xi_3 I_{11}
\eeqa
 where $k$ becomes  off-shell gauge field's momentum and $I_{10},I_{11}$ are
 \beqa
I_{10}&=& (-k_1.k_2 k_3.k\xi_{1a}+k_1.k_2 \xi_1.k k_{3a}+\xi_1.k_2 k_3.k k_{1a}-k_2.\xi_1k_1.k k_{3a})\nonumber\\
 I_{11}&=&(-k.k_2 k_3.k_1\xi_{1a}+k.k_2 \xi_1.k_3 k_{1a}+\xi_1.k k_3.k_1 k_{2a}-k_3.\xi_1k_1.k k_{2a})
 \eeqa
 Now by applying the hermition conjugate of $\Tr\left(a_{n,m}\cD_{nm}[D_a \phi^i D^b \phi_i F^{ac}F_{bc}]\right)$ we are lead to
 \beqa
a_{n,m}\xi_2.\xi_3 \bigg((k.k_3)^n(k_1.k_3)^m  I_{10}+
(k.k_3)^m(k_1.k_3)^n  I_{11}\bigg)
\eeqa

  Therefore one must do careful computations for all the other couplings in \reef{highder} and also should consider their hermition conjugate as well. The final result is
  \beqa
   V_{\beta}^{b}(A,\phi_2,\phi_3,A_1)&=&\frac{T_{p}}{2}\xi_2.\xi_3 \frac{1}{2\pi^2}(\alpha')^{n+m}(a_{n,m}+b_{n,m})
\bigg(\frac{}{}(k_2\inn k)^m(k_1\inn k_2)^n+(k_2\inn k)^n(k_2\inn k_1)^m
\nonumber\\&&+(k_1\inn k_3)^m(k_3\inn k)^n+(k\inn k_3)^m (k_1\inn k_3)^n
\bigg)  (2\pi\alpha')^4 \quad \Tr(\lam_1\lam_2\lam_3\lambda_{\beta})\nonumber\\&&\times
\bigg[\xi_{1b}ts- 2tk_3.\xi_{1}k_{2b}
-2sk_2.\xi_{1}k_{3b}\bigg],\labell{verppaa2}\eeqa

Having set \reef{verppaa2},\reef{Fey2}  into \reef{amp5491}, we get the infinite massless gauge field poles of the amplitude in field theory side:
  \beqa
&&-32\pi\mu_p\frac{\eps^{a_{0}\cdots a_{p-1}a}\xi_2.\xi_3
H^{(p)}_{a_0\cdots a_{p-1}}}{(p)!(s+t+u)}\Tr(\lam_1\lam_2\lam_3)
\sum_{n,m=0}^{\infty}\bigg((a_{n,m}+b_{n,m})[s^{m}t^{n}+s^{n}t^{m}]\nonumber\\&&
\bigg[\xi_{1a}ts- 2tk_3.\xi_{1}k_{2a}
-2sk_2.\xi_{1}k_{3a}\bigg]
\label{amphigh8}\eeqa

In order to check the field theory amplitude with string amplitude \reef{nb2} one needs to actually have some of the coefficients such as
\beqa
&&a_{0,0}=-\frac{\pi^2}{6},\,b_{0,0}=-\frac{\pi^2}{12},a_{1,0}=2\z(3),\,a_{0,1}=0,\,b_{0,1}=-\z(3),a_{1,1}=a_{0,2}=-7\pi^4/90,\nonumber\\
&&a_{2,2}=(-83\pi^6-7560\z(3)^2)/945,b_{2,2}=-(23\pi^6-15120\z(3)^2)/1890,a_{1,3}=-62\pi^6/945,\label{mmnnb}\nonumber\\
&&\,a_{2,0}=-4\pi^4/90,\,b_{1,1}=-\pi^4/180,\,b_{0,2}=-\pi^4/45,a_{0,4}=-31\pi^6/945,a_{4,0}=-16\pi^6/945,\nonumber\\
&&a_{1,2}=a_{2,1}=8\z(5)+4\pi^2\z(3)/3,\,a_{0,3}=0,\,a_{3,0}=8\z(5),b_{1,3}=-(12\pi^6-7560\z(3)^2)/1890,\nonumber\\
&&a_{3,1}=(-52\pi^6-7560\z(3)^2)/945, b_{0,3}=-4\z(5),\,b_{1,2}=-8\z(5)+2\pi^2\z(3)/3,\nonumber\\
&&b_{0,4}=-16\pi^6/1890.\eeqa

Notice the fact that $b_{n,m}$ must be symmetric and concerning T-duality transformation these coefficients are the same as those have been appeared  for one RR and 3 gauge fields \cite{Hatefi:2010ik}. Later on we will go through all of the contact terms for $p=n$ case, even those terms which have been cancelled out with the resulted propagator in the above field theory amplitude. Although the method for obtaining them with all needed details have been explained in \cite{Hatefi:2012ve,Hatefi:2012wj}.

\vskip 0.1in

Meanwhile the amplitude in string theory is given in \reef{nb2}. If  the higher derivative couplings of \reef{highder} are correct, we must be able to produce exactly all massless poles in \reef{nb2}. To do so, first we omit similar coefficients from both string and field amplitudes and then compare \reef{amphigh8} with \reef{nb2} order by order. In the other words, the aim is to compare
\beqa
-\mu_p \pi\sum_{n,m=0}^{\infty}\bigg((a_{n,m}+b_{n,m})[s^{m}t^{n}+s^{n}t^{m}]\bigg)
\eeqa
with
\beqa
2^{-1}\pi^3 \mu_p \sum_{n,m=0}^{\infty}c_{n,m}\bigg ( s^{m}t^{n}+s^{n}t^{m}\bigg)
\eeqa
 By applying   $n=m=0$, at zeroth order of $\alpha'$ we get
\beqa
-2\pi(a_{0,0}+b_{0,0})&=&-2\pi(\frac{-\pi^2}{6}+\frac{-\pi^2}{12})=\frac{\pi^3}{2} (2 c_{0,0})\eeqa
 At first order of $\alpha'$, we find
\beqa
-\pi(a_{1,0}+a_{0,1}+b_{1,0}+b_{0,1})(s+t)&=&0=\frac{\pi^3}{2} ( c_{1,0}+c_{0,1})(s+t)\nonumber\eeqa
  At the  second order of $(\alpha')$, we lead to
\beqa
&&-2\pi(a_{1,1}+b_{1,1})st-\pi(a_{0,2}+a_{2,0}+b_{0,2}+b_{2,0})[s^2+t^2]\nonumber\\
&&=\frac{\pi^5}{6}(st)+\frac{\pi^5}{6}(s^2+t^2)\nonumber\\&&=\frac{\pi^3}{2}[c_{1,1}(2st)+(c_{2,0}+c_{0,2})(s^2+t^2)]
\nonumber\eeqa
At third order of $\alpha'$, we gain
\beqa
&&- \pi(a_{3,0}+a_{0,3}+b_{0,3}+b_{3,0})[s^3+t^3]- \pi(a_{1,2}+a_{2,1}+b_{1,2}+b_{2,1})[st(s+t)]\nonumber\\
&&=-4\pi^3\xi(3)st(s+t)=\frac{\pi^3}{2}[(c_{0,3}+c_{3,0})[s^3+t^3]+(c_{2,1}+c_{1,2})st(s+t)]
\nonumber\eeqa
In order to be sure we have obtained the correct couplings with exact coefficients, we want to go ahead one more order so at fourth order of $(\alpha')$, we find the following numerical factor
\beqa
&&- \pi(a_{4,0}+a_{0,4}+b_{0,4}+b_{4,0})(s^4+t^4)-\pi (a_{3,1}+a_{1,3}+b_{3,1}+b_{1,3})[st(s^2+t^2)]\nonumber\\
&&-2\pi(a_{2,2}+b_{2,2}) s^2t^2=\frac{\pi^7}{15}(s^4+t^4+2(s^3t+t^3s)+3s^2t^2)\nonumber\\&&
=\frac{\pi^3}{2}[(c_{4,0}+c_{0,4})(s^4+t^4)+(c_{1,3}+c_{3,1})(s^3t+t^3s)+2c_{2,2} s^2t^2]\nonumber\eeqa

We have  highly used the coefficients in \reef{mmnnb}.
In general all checks to all orders in $\alpha'$ can be carried out to indeed see that all  massless gauge poles of $<V_CV_AV_{\phi} V_{\phi}>$
are produced.
Therefore we come to important fact that these couplings do work out even for the amplitude of $CA\phi\phi$, and this is the other recent point that comes out from our attempts which has been hidden  in \cite{Garousi:2000ea} for a while.
 Therefore not only  does it confirm that our recent higher derivative couplings are exact  up to on-shell ambiguity but also it resolves the fact that
$p_ap^a$ must tend to zero to get the correct expansion  for all BPS branes.


\subsection{Infinite massless $t,s$-channel scalar  poles for $p+2=n$ case }

The goal in this section is to actually produce all infinite  s-channel and t-channel scalar poles. The first simple scalar  pole in t-channel has already been produced in \cite{Garousi:2000ea} but again in there  all infinite scalar poles have been overlooked, however we are going to come over them as well.

\vskip 0.1in

Having taken our recent ideas for Super Yang-Mills \cite{Hatefi:2010ik,Hatefi:2012ve,Hatefi:2012rx}, we show that the same arguments here also hold.
By applying $us L_2', tu L_2'$ expansions into all terms (except the second term) in the ${\cal A}_{2}$ amplitude    and extracting the traces, one can find out all massless scalar  poles in t  channel  for string amplitude as

 \beqa
 &&\frac{-16\pi^2 \mu_p}{(p+1)!}
\bigg\{-2(p+1)\xi_{1}.k_{2}\xi_{2i}k_{3c}\xi_{3j}\eps^{a_{0}\cdots a_{p-1}c}H^{ij}_{a_{0}\cdots a_{p-1}}
+2 \xi_{2i}p^j\xi_1.k_2\xi_{3j}\eps^{a_{0}\cdots a_{p}}H^{i}_{a_{0}\cdots a_{p}}
\bigg\}\nonumber\\&&\times\sum_{n=-1}^{\infty}b_n\bigg(\frac{1}{t}(u+s)^{n+1}\bigg)\Tr(\lam_1\lam_2\lam_3)
\label{tpoles}\eeqa

 All s-channel poles are also written down as below

\beqa
  &&\frac{-16\pi^2 \mu_p}{(p+1)!}
\bigg\{2(p+1) k_{3}.\xi_{1}k_{2b}\xi_{3j}\xi_{2i}\eps^{a_{0}\cdots a_{p-1}b}H^{ji}_{a_{0}\cdots a_{p-1}}
-2 \xi_{3j}p^i\xi_1.k_3\xi_{2i}\eps^{a_{0}\cdots a_{p}}H^{j}_{a_{0}\cdots a_{p}}\bigg\}\nonumber\\&&\times\sum_{n=-1}^{\infty}b_n\bigg(\frac{1}{s}(u+t)^{n+1}\bigg)\Tr(\lam_1\lam_2\lam_3)
\label{spoles}\eeqa

By interchanging scalars in the ${\cal A}_{2}$ amplitude, we reach to the point that the amplitude is anti symmetric thus in order to make sense of our computations one has to consider non-Abelian gauge group. As it is clear from \reef{spoles} and \reef{tpoles}, once we produced all massless  t-channel scalar poles,  all infinite s-channel scalar poles can be easily produced by replacing $s\leftrightarrow t$ and re-labeling $2\leftrightarrow 3$ in all their momenta and polarizations.
Therefore let us just produce all infinite massless scalar poles in t-channel in field theory.

The Feynman rule in field theory to produce all t-channel poles should be followed by

\beqa
{\cal A}&=&V^i_{\alpha}(C_{p+1},\phi_3,\phi)G^{ij}_{\alpha\beta}(\phi)V^j_{\beta}(\phi,A_1,\phi_2),\labell{amp443}
\eeqa

such that the vertex of $V^j_{\beta}(\phi,A_1,\phi_2)$ should be found from the scalar field's kinetic term like $\frac{(2\pi\alpha')^2}{2}  \Tr(D_a\phi^i D^a\phi_i)$
where all possible orderings must be regarded in field theory as well. Therefore

\beqa
V^j_{\beta}(\phi,A_1,\phi_2)&=&-2i\lambda^2T_p k_2.\xi_1
\xi^j_2 \Tr(\lambda_1\lambda_2\lambda_\beta)
\nonumber\\
G_{\alpha\beta}^{ij}(\phi)&=&\frac{-i}{N\lambda^2T_p}\frac{\delta^{ij}
\delta_{\alpha\beta}}{k^2}\,\,\, ,
\label{ver13}
\eeqa

 $k^2=-(k_2+k_1)^2=t$  should be substituted  in the propagator.
As argued in the last section scalar field 's kinetic term has been fixed so it has no correction, the simple scalar t-channel pole
has no correction either. Therefore not only we need to find $V^i_{\alpha}(C_{p+1},\phi_3,\phi)$ but also its higher derivative corrections are also needed.

First of all let us discuss how to produce $V^i_{\alpha}(C_{p+1},\phi_3,\phi)$ without taking its higher derivative.

The first coupling between one gauge field, two scalar fields and one RR should be included from Myers ' terms. Namely, we may think of the coupling between a  commutator of transverse scalars and a world volume field strength of gauge field and one RR -$(p+1)$ form field as we call it $S_6$

\beqa
S_{6}&=&{i\over4}(2\pi\alpha')^2\mu_p\int d^{p+1}\sigma {1\over(p-1)!} \eps^{a_0\cdots a_{p}}
\,\Tr\left(F_{a_0a_1}[\Phi^j,\Phi^i]\right)
C^{(p+1)}_{ija_2\cdots a_{p}} .
\label{33}
\eeqa
For more details on Chern-Simons actions, Taylor expansion and Pull-back , one should deal with section 5 of \cite{Hatefi:2012wj}.
In addition to \reef{33} we need to know two more couplings, basically first we need to use Taylor expansion very properly for this case as
\beqa
S_{7}&=&{(2\pi\alpha')^2\mu_p\over2}\int d^{p+1}\sigma {1\over (p+1)!}\eps^{a_0\cdots a_{p}}
  \Tr\left(\Phi^j \Phi^i\right)
\prt_j\prt_iC^{(p+1)}_{a_0\cdots a_{p}}\nonumber\\&&
={(2\pi\alpha')^2\mu_p\over2}\int d^{p+1}\sigma {1\over (p+1)!}
\eps^{a_0\cdots a_{p}} \Tr\left(\Phi^j\Phi^i\right)
\prt_jH^{(p+2)}_{ia_0\cdots a_{p}}
\label{221}
\eeqa
such that $H^{p+2}=dC^{p+1}$, the other couplings which are vital for our case must be read from Pull-back ,namely we shall point out to the following couplings as well

\beqa
S_{8}
&=&\frac{(2\pi\alpha')^2\mu_p}{2}\int d^{p+1}\sigma {1\over (p+1)!}
\eps^{a_0\cdots a_{p}}\bigg[p(p+1)\,
\Tr\left(D_{a_0}\Phi^i\,D_{a_1}\Phi^j\right)
C^{(p+1)}_{ija_2\cdots a_{p}}\nonumber\\&&+2(p+1)
\Tr\left(\Phi^j D_{a_0}\Phi^i\right)
\prt_jC^{(p+1)}_{ia_1\cdots a_{p}}
\bigg]
\nonumber
\eeqa
 Having taken integration by parts and adding some of the actions we reach to

\beqa
S_{6}+S_{8}&=&{(2\pi\alpha')^2\over2}\mu_p\int d^{p+1}\sigma {1\over (p+1)!}
\eps^{a_0\cdots a_{p}}\left[(p+1)\Tr\left(D_{a_0}\Phi^j\Phi^i\right)
H^{(p+2)}_{ija_1\cdots a_{p}}\right]
\nonumber\eeqa

In order to get several contributions , one has to extract the covariant derivative of scalar field $(D_a\phi^i=\partial_a\phi^i+i[A_a,\phi^i])$ such that
\beqa
S_{6}+S_{8}&=&{(2\pi\alpha')^2\over2}\mu_p\int d^{p+1}\sigma {1\over (p+1)!}
\eps^{a_0\cdots a_{p}}\bigg[2i(p+1)\Tr\left(A_{a_0}\Phi^j\Phi^i\right)
H^{(p+2)}_{ija_1\cdots a_{p}}\nonumber\\&&+(p+1)\Tr\left(\prt_{a_0}\Phi^j\Phi^i\right)
H^{(p+2)}_{ija_1\cdots a_{p}}\bigg]
\label{esi23}
\eeqa

 Note that the first term  in \reef{esi23} will be employed in the next section to obtain all the contact terms $p+2=n$ case for three open strings, namely two scalars, one gauge field and one closed string RR-$p+1$ form field.

However, in order to obtain  $V^i_{\alpha}(C_{p+1},\phi_3,\phi)$ for producing the first massless scalar pole,one must add the relevant couplings together at leading order , basically by adding some of the couplings as below

\beqa
\frac{\mu_p(2\pi\alpha')^2}{2(p+1)!}\int d^{p+1}\sigma
\eps^{a_0\cdots a_{p}}\left[\Tr\bigg(\Phi^j\Phi^i\bigg)
\prt_j H^{(p+2)}_{ia_0\cdots a_{p}}
+(p+1)\Tr\left(\prt_{a_0}\Phi^j\Phi^i\right)
H^{(p+2)}_{ija_1\cdots a_{p}}\right]
\label{5ghtt}
\eeqa

we can easily get the leading vertex of $V^i_{\alpha}(C_{p+1},\phi_3,\phi)$ as
\beqa
V^i_{\alpha}(C_{p+1},\phi_3,\phi)&=&\frac{N\mu_p(2\pi\alpha')^2}{(p+1)!}\Tr(\lambda_3\lambda_{\alpha})
\eps^{a_0\cdots a_{p}}\bigg[p^j \xi_{3j}H^{i}_{a_0\cdots a_{p}}\nonumber\\&&
+(p+1)H^{ij}_{a_1\cdots a_{p}}k_{3a_{0}} \xi_{3j} \bigg]
\label{ppo}\eeqa
 Needless to say that $N$ is indeed the normalisation constant to be chosen for all U(N) generators, such that
 \beqa
\xi_{1i}&=&\xi_{1i}^{\alpha} Q_{\alpha},\quad\quad\quad N\delta^{\alpha\beta}=\Tr(Q^\alpha Q^\beta) \eeqa
Now by replacing \reef{ppo} and \reef{ver13} into \reef{amp443}, we are able to just produce exactly the first simple t-channel pole in \reef{tpoles}. In order to produce all infinite t-channel  poles, we should look for all higher derivative corrections of \reef{5ghtt}. One can apply the main ideas of \cite{Hatefi:2010ik, Hatefi:2012ve} to indeed get the all higher derivative corrections of \reef{5ghtt} as

\beqa
&&\frac{\mu_p(2\pi\alpha')^2}{2(p+1)!}\int d^{p+1}\sigma
\eps^{a_0\cdots a_{p}} \sum_{n=-1}^{\infty} b_n (\alpha')^n \bigg[\Tr\bigg(D_{a_{1}...a_{n}}\Phi^jD^{a_{1}...a_{n}}\Phi^i\bigg)
\prt_j H^{(p+2)}_{ia_0\cdots a_{p}}\nonumber\\&&
+(p+1)\Tr\left(\prt_{a_0}D_{a_{1}...a_{n}}\Phi^jD^{a_{1}...a_{n}}\Phi^i\right)
H^{(p+2)}_{ija_1\cdots a_{p}}\bigg]
\label{5gh}
\eeqa
The important point here is that the commutators in covariant derivative of scalar fields do not play any role and in fact they have no contribution to above vertex so all covariant derivatives can be replaced with their own partial derivatives.

  By constructing the correct higher derivative corrections of \reef{5ghtt} as \reef{5gh}, one can write down the general form of the needed vertex as

  \beqa
V^i_{\alpha}(C_{p+1},\phi_3,\phi)&=&\frac{N\mu_p(2\pi\alpha')^2}{(p+1)!}\Tr(\lambda_3\lambda_{\alpha})
\eps^{a_0\cdots a_{p}}\sum_{n=-1}^{\infty} b_n (\alpha'k_3.k)^n\bigg[p^j \xi_{3j}H^{i}_{a_0\cdots a_{p}}\nonumber\\&&
+(p+1)H^{ij}_{a_1\cdots a_{p}}k_{3a_{0}} \xi_{3j} \bigg]
\label{ppo2}\eeqa
   where $\sum_{n=-1}^{\infty} b_n (\alpha'k_3.k)^n=\sum_{n=-1}^{\infty} b_n (s+u)^n$ has been used.

Therefore by making use of  this new higher vertex (to all orders of $\alpha'$) \reef{ppo2} and substituting
  \reef{ppo2} and \reef{ver13} into \reef{amp443}, fortunately we were able to exactly obtain all infinite t-channel scalar poles in \reef{tpoles} as follows

 \beqa
 {\cal A}&=&\frac{-16\pi^2 \mu_p}{(p+1)!}
\bigg\{-2(p+1)\xi_{1}.k_{2}\xi_{2i}k_{3a}\xi_{3j}\eps^{a_{0}\cdots a_{p-1}a}H^{ij}_{a_{0}\cdots a_{p-1}}
+2 \xi_{2i}p^j\xi_1.k_2\xi_{3j}\eps^{a_{0}\cdots a_{p}}H^{i}_{a_{0}\cdots a_{p}}
\bigg\}\nonumber\\&&\times\sum_{n=-1}^{\infty}b_n\bigg(\frac{1}{t}(u+s)^{n+1}\bigg)\Tr(\lam_1\lam_2\lam_3)
\label{tpoles2}\eeqa

Having replaced $t\leftrightarrow s$ and  $2\leftrightarrow 3$ we can also produce all infinite s-channel scalar poles as well.

Thus up to pole levels we observe that field theory does agree with string amplitude, however in the next sections we will see that there are some contact terms in string theory such that their field theory is unknown. It is remarkable to note that  these sort of new interactions neither can be found by Myers'terms nor with Taylor/Pull back method. To our knowledge pull-back should be corrected \cite{Hatefi:2012ve}. Essentially we find some new couplings by comparing them with the exact result of string amplitude. After carrying out long computations and producing all infinite massless scalar, gauge poles for all possible different channels , let us go further and talk about contact interactions and new Wess-Zumino couplings, which can be found just by direct S-Matrix computations.

\section{Contact interactions for $p=n$ case}

 Notice the fact that $p=n$ does mean  that we are taking into account all $C_{p-1}$ couplings to $D_{(p-2)}$-brane.
By taking $n=p$ case  , the final form of our amplitude
reduced to the following interactions:
\beqa
A^{CA\phi\phi}&=&\frac{i(2\pi\alpha')^3\mu_p}{2p!}\eps^{a_0\cdots a_{p-1}a} \xi_{1a}\xi_{2i}\xi_{3j}\bigg(
 p^{i}p^{j}H_{a_0\cdots a_{p-1}}+p(p-1)k_{3a_0}k_{2a_1}
H^{ij}_{a_2\cdots a_{p-1}}
 \nonumber\\
&&-p k_{2a_0}p^jH^i_{a_1\cdots a_{p-1}}-
p k_{3a_0}p^iH^j_{a_1\cdots a_{p-1}}\bigg)
\nonumber\\&&
\times\bigg( \sum_{n=0}^{\infty}c_n(s+t+u)^{n+1}+\sum_{n,m=0}^{\infty}c_{n,m}[(s)^n(t)^m +(s)^m(t)^n]\nonumber\\&&
+\sum_{p,n,m=0}^{\infty}f_{p,n,m}(s+t+u)^{p+1}[(s+t)^n(st)^{m}]\bigg),
 \label{cterm2}
\eeqa
Some of the contact terms just at the leading order for this  case were known in \cite{Garousi:2000ea}. Note that we are just considering $\Tr(\lambda_1\lambda_2\lambda_3)$ while the amplitude has the other possible ordering which is $\Tr(\lambda_1\lambda_3\lambda_2)$. In order to obtain full amplitude one has to replace $s\leftrightarrow u$ and $2\leftrightarrow 3$  in the above contact interactions and add them up with \reef{cterm2}. In order to produce  \reef{cterm2} we have to consider several couplings from field theory.
Let us first reconsider the couplings between two scalars (coming from pull-back) and one gauge field  as

 \beqa
S_{9}&=& \frac{(2\pi\alpha')^3\mu_p}{4(p-1)!}\int d^{p+1}\sigma \eps^{a_0\cdots a_p}
\bigg( (p-2)(p-1)\STr(F_{a_0a_1}D_{a_2}
\Phi^i D_{a_3}\Phi^j)C_{ija_4\cdots a_p}
\nonumber\\&&+2(p-1) \STr(F_{a_0a_1} D_{a_2}\Phi^i\Phi^j)\,\prt_jC_{ia_3\cdots a_p}\bigg)
\labell{cc454}
\eeqa

The important point for the above coupling is that commutator in the definition of
covariant derivative of scalars must be overlooked as we are looking for two scalars and one gauge coupling.
As a matter of fact all covariant derivative should be replaced with their partial derivative.

\vskip 0.1in

The other coupling (which is essential for producing all contact terms for $p=n$  case), should be coming from Taylor expansions as follows
\beqa
S_{10}&=& \frac{(2\pi\alpha')^3\mu_p}{4(p-1)!}\int d^{p+1}\sigma \eps^{a_0\cdots a_p}
\STr (F_{a_0a_1}
\Phi^i\Phi^j) \prt_i\prt_jC_{a_2\cdots a_p}
\labell{rr}
\eeqa

Now by extracting field strength, replacing all covariant derivatives with their partial derivatives, adding
\reef{rr} and \reef{cc454} $(S_{11}=S_{9}+S_{10})$ and finally replacing all fields to their polarizations and in particular  changing derivatives to momenta , we can precisely produce all contact terms at the leading order with the following couplings :

\beqa
S_{11}&=&\frac{\lambda^3\mu_p}{2p!}\int d^{p+1}\sigma \eps^{a_0\cdots a_p}\bigg(
p(p-1)\STr(A_{a_0}\prt_{a_1}\Phi^i\prt_{a_2}\Phi^j)
H_{ija_3\cdots a_p}\nonumber\\&&
+2p \STr(A_{a_0}\prt_{a_1}\Phi^j\Phi^i)\,\prt_i H_{ja_2\cdots a_p}
+\STr (A_{a_0}\Phi^i\Phi^j)
\prt_i\prt_j H_{a_1\cdots a_p}\bigg)
\labell{nnv}
\eeqa

Note that the first term in \reef{nnv} is derived from \reef{cc454} where all the commutators should be dropped, as we are looking for the couplings between two scalars and one gauge and one RR -$p-1$ form field.

\vskip 0.1in

Also note that, in the first term of \reef{cc454}, the partial derivative inside the field strength can not act on scalars because the $\eps$ tensor is antisymmetric
and the multiplication of symmetric tensor  and antisymmetrc tensor  becomes zero so it can act just on RR field.
Symmetric trace does mean  that, taking average on the whole possible orderings of the fields is vital. The appearance of symmetric trace for the last term is necessary as we have to produce  the third and the last term in \reef{cterm2} very precisely.

In order to produce all infinite contact interactions in \reef{cterm2}, the following higher  derivative  corrections should have been taken into account.
\beqa
(st)^{m}HA\phi\phi&=&(\alpha')^{2m}H \partial_{a_1}\cdots \partial_{a_{2m}}A D^{a_{1}}\cdots D^{a_{m}}\phi
D^{a_{m+1}}\cdots D^{a_{2m}}\phi,\nonumber\\
(s+t)^{n}HA\phi\phi&=&(\alpha')^{n}H  \partial_{a_1}\cdots \partial_{a_{n}} A D^{a_{1}}\cdots D^{a_{n}}(\phi
\phi),\nonumber\\
(s)^{m}t^n HA\phi\phi&=&(\alpha')^{n+m}H \partial_{a_1}\cdots \partial_{a_{n}} \partial_{a_{1}}\cdots \partial_{a_{m}}A D^{a_{1}}\cdots D^{a_{n}}\phi D^{a_{1}}\cdots D^{a_{m}}\phi,
\nonumber\\
(s)^{n}t^m HA\phi\phi&=&(\alpha')^{n+m}H \partial_{a_1}\cdots \partial_{a_{n}} \partial_{a_{1}}\cdots \partial_{a_{m}}A D^{a_{1}}\cdots D^{a_{m}}\phi D^{a_{1}}\cdots D^{a_{n}}\phi,
\nonumber\\
(s+t+u)^{p+1} HA\phi\phi&=&(\frac{\alpha'}{2})^{p+1}H (D_{a}D^{a})^{p+1}(A\phi\phi).
\labell{67s282ee}
\eeqa

The important point which must be highlighted is that, the connection part or the commutator in the covariant derivative of scalars must be dropped in the above couplings.

\section{New couplings  for $p=n$ case}

\vskip 0.1in

Carrying out the trace and considering only all related contact interactions for  $A_{3}$, we get the terms like
\beqa
A_{3}^{CA\phi\phi  }&=&-\frac{16 \pi^3\mu_p}{(p)!}\xi_3.\xi_2
H_{a_0\cdots a_{p-1}}\eps^{a_0\cdots a_{p-1}a}
[\frac {ts}{2}\xi_{1a}-2tk_3.\xi_1 k_{2a}+2\leftrightarrow 3]\nonumber\\&& \times\bigg( \sum_{n=0}^{\infty}c_n(s+t+u)^{n}
+\sum_{p,n,m=0}^{\infty}f_{p,n,m}(s+t+u)^{p} (st)^{m}(s+t)^{n}\bigg)
\label{case32}\eeqa
These new contact terms do satisfy related Ward identity .
It is really worth trying to point out that these   contact terms  are sort of new couplings. In principle we should be able to
  produce these new couplings by introducing new couplings in field theory .

Keep in  mind that, we  proceed to find out new couplings in field theory term by term however after all, one has to add all of them together.
 Let us first produce the first term in \reef{case32} at leading order (remember $c_1=\frac{\pi^2}{6}, f_{0,1,0}=\frac{-\pi^2}{3}$)
  then we generalize  its higher order corrections to all orders in $\alpha'$, namely consider
\beqa
-\frac{16 \pi^3\mu_p}{(p)!}\xi_3.\xi_2
H_{a_0\cdots a_{p-1}}\eps^{a_0\cdots a_{p-1}a}\frac {ts}{2}\xi_{1a}\bigg( \frac{\pi^2}{6}(s+t+u)
-\frac{\pi^2}{3}(s+t)\bigg)
\label{case322}\eeqa

One can exactly produce \reef{case322} by taking into account the fact that the sum of world volume indices must cover all world volume indices such that
the first term in \reef{case322} is obtained by the new coupling  as
\beqa
S_{12}&=&\frac{(2\pi\alpha')\mu_p (\pi^2\alpha')^2}{3}\int d^{p+1}\sigma\ C_{(p-1)}\wedge \bigg(\frac{\alpha'}{2}(D^cD_c)(D^bD^aF D_a\Phi^iD_b\Phi_i) \bigg),
\labell{snjh2221}
\eeqa

Now we can generalize the above coupling  \reef{snjh2221} to produce all non leading couplings for the first term in \reef{case322} as
\beqa
S_{13}&=& \frac{(2\pi\alpha')^3\mu_p}{2}  \int d^{p+1}\sigma \sum_{n=0}^{\infty}c_n (\frac{\alpha'}{2})^n C_{(p-1)}
\wedge (D^cD_c)^n(D^bD^aF D_a\Phi^iD_b\Phi_i),
\labell{snjh1122}
\eeqa

In those couplings, we have written explicit covariant derivative of the scalar fields , however with our computations we can confirm the presence of
just the partial derivatives in covariant derivatives, thus in order to check whether or not commutators should be held, one should perform  higher point functions just like $CAA\phi\phi$ \cite{Hatefi:2013eh}.

\vskip 0.1in

Pursuing the argument mentioned above, one can produce the second term in \reef{case322}
as follows

\beqa
S_{14}&=&-\frac{(8\pi\alpha')\mu_p (\alpha'\pi^2)^2}{3}\int d^{p+1}\sigma\ C_{(p-1)}\wedge (D^cD^aD^bF D_c[D_a\Phi^i D_b\Phi_i]),
\labell{snjh22}
\eeqa

Making use of the steps have been mentioned in \cite{Hatefi:2012wj}, we are able to
 get the all higher order corrections for the second term in \reef{case322} as

\beqa
S_{15}&=&-\frac{(2\pi\alpha')^3\mu_p }{2}\int d^{p+1}\sigma     \sum_{p,n,m=0}^{\infty}f_{p,n,m}(\frac{\alpha'}{2})^p (\alpha')^{2m+n}
C_{(p-1)}\wedge (D^aD_a)^p (D_{b_1}\cdots D_{b_{2m}} \nonumber\\&&D^{a_1}\cdots D^{a_n}  D^bD^cF D_{a_1}\cdots D_{a_n}[D^{b_1}
\cdots D^{b_m}D_b\Phi^iD^{b_{m+1}}\cdots D^{b_{2m}}D_c\Phi_i]),
\labell{snjh229}
\eeqa

 In order to produce  the rest of the terms in \reef{case32}, namely one has to consider the following terms
\beqa
&&-\frac{16 \pi^3\mu_p}{(p)!}\xi_3.\xi_2
H_{a_0\cdots a_{p-1}}\eps^{a_0\cdots a_{p-1}a}
[-2tk_3.\xi_1 k_{2a}+2\leftrightarrow 3]\nonumber\\&& \times\bigg( \sum_{n=0}^{\infty}c_n(s+t+u)^{n}
+\sum_{p,n,m=0}^{\infty}f_{p,n,m}(s+t+u)^{p} (st)^{m}(s+t)^{n}\bigg)
\label{case32yin}\eeqa

To produce the first term  in \reef{case32yin}  at leading order, one must consider  the following coupling  and adds it up to \reef{snjh2221} and \reef{snjh22}

 \beqa
S_{16}&=&\frac{(2\pi\alpha')^3\mu_p \pi^2 \eps^{a_0 \cdots a_{p}}H_{a_0 \cdots a_{p-1}}\alpha' (D_a D^a)}{12p!} \int d^{p+1}\sigma
 \bigg(\partial_{c} A_b  \partial_{a_p} D^c\Phi_i D^b \Phi^i -2\leftrightarrow 3\bigg)
\labell{vv}
\eeqa

Now we can generalize the above coupling  to produce all non leading couplings to all orders of $\alpha'$
\beqa
S_{17}&=&\frac{(2\pi\alpha')^3\mu_p  }{p!} \eps^{a_0 \cdots a_{p}}\int d^{p+1}\sigma
H_{a_0 \cdots a_{p-1}} \sum_{n=0}^{\infty}c_n (\frac{\alpha'}{2})^n (D_a D^a)^n \nonumber\\&&\times
\bigg(\partial_{c} A_b  \partial_{a_p} D^c\Phi_i D^b \Phi^i
-2\leftrightarrow 3\bigg)
\labell{bbcx}
\eeqa

Also notice that the second term in \reef{case32yin} at leading order can be reproduced as

\beqa
S_{18}&=&\frac{(2\pi\alpha')^3\mu_p \pi^2 }{3p!} \eps^{a_0 \cdots a_{p}}\int d^{p+1}\sigma
H_{a_0 \cdots a_{p-1}}  \bigg(D^d\partial_{c} A_b  D_d [\partial_{a_p} D^c\Phi_i D^b \Phi^i ]-2\leftrightarrow 3 \bigg)
\labell{vv223}
\eeqa

To find  all higher orders in $\alpha'$ one must take into account the derivatives as appeared in \reef{snjh229} such that the final form is

\beqa
S_{19}&=&\frac{(2\pi\alpha')^3\mu_p \eps^{a_0 \cdots a_{p}}H_{a_0 \cdots a_{p-1}}}{p!} \int d^{p+1}\sigma
\sum_{p,n,m=0}^{\infty}f_{p,n,m}(\frac{\alpha'}{2})^p (\alpha')^{2m+n}
 (D^aD_a)^p
\bigg( D_{b_1}\cdots D_{b_{2m}}\nonumber\\&&\times D_{a_1}\cdots D_{a_n} \partial_{c}A_b  D^{a_1}\cdots D^{a_n}
[\partial_{a_p} D^{b_1}
\cdots D^{b_m} D^c\Phi_i D^b D^{b_{m+1}}\cdots D^{b_{2m}}\Phi^i ] -2\leftrightarrow 3\bigg)
\nonumber\\&&
\labell{vv22r6io}
\eeqa

Of course for our amplitude $(CA\phi\phi)$ we just could confirm the presence of partial derivatives in the definitions of the covariant derivatives  in the coupling \reef{vv} and it remains an open question to check whether or not the commutator in the definitions of covariant derivative of scalar fields will be kept. In order to answer this subtlety one must perform higher point functions, namely to compute either $CAA\phi\phi$ or  $CAAA\phi\phi$. However there are some subtleties to answer this question  and some of them have been addressed in \cite{Hatefi:2010ik,Hatefi:2012wj}. We hope to answer some of them in future
\cite{Hatefi:2013eh}.

\vskip 0.2in

Note that these couplings are consistent with string amplitude of one $(p-1)$-form closed string RR and two scalar and one gauge field and they are  new in the sense that neither do they come from Myers' terms, Pull-back, Taylor expansion nor expanding the exponential and suing the multiplication rule of the super matrices.

 \vskip 0.1in

As it stands and it is known , applying the direct computations of string amplitudes is the only consistent and a quite reasonable method to actually find out  new couplings in field theory.


Thus to conclude, in order to produce exactly the contact interactions in \reef{case32} at leading order one has to consider the sum of $S_{12},S_{14},S_{16},S_{18}$ and consider these interactions by replacing $2\leftrightarrow 3$ as well.

\vskip 0.2in

Finally we get to those contact terms that  have been overlooked in  section 4.
 By making use of some valuable formula we might write those interactions down as

\beqa
\frac{16\pi\mu_p}{p!} \eps^{a_{0}\cdots a_{p-1}a} H_{a_{0}\cdots a_{p-1}}\xi_3.\xi_2\bigg[\xi_{1a}ts-2tk_3.\xi_{1}k_{2a}
-2sk_2.\xi_{1}k_{3a}\bigg]
\sum_{n,m=0}^{\infty}(a_{n,m}+b_{n,m})(-\alpha' k^2)^{l-1}\nonumber\\
\left[\left(2\sum_{l=1}^m \pmatrix{m\cr
l}(s^{m-l}t^n+t^{m-l}s^n)+2\sum_{l=1}^n \pmatrix{n\cr
l}(s^{n-l}t^m+t^{n-l}s^m)\right) \right.\nonumber\\
\left.+\sum_{l=1,j=1}^{n,m} \pmatrix{n\cr
l}\pmatrix{m\cr j}(s^{n-l}t^{m-j}+t^{n-l}s^{m-j})(-\alpha'k^2)^{j}\right]
\Tr(\lam_1\lam_2\lam_3)\nonumber\eeqa

The important point is that we can also write these contact terms in a closed form as follows
\beqa
 \frac{16\pi\mu_p}{p!} \eps^{a_{0}\cdots a_{p-1}a} H_{a_{0}\cdots a_{p-1}}\xi_3.\xi_2\bigg[\xi_{1a}ts-2tk_3.\xi_{1}k_{2a}
-2sk_2.\xi_{1}k_{3a}\bigg]\nonumber\\\times
 \Tr(\lam_1\lam_2\lam_3)
\sum_{p,n,m=0}^{\infty}f'_{p,n,m}(s+t+u)^p(s+t)^n(st)^{m},\label{uuyy}\eeqa

We should have pointed out the fact that  $f'_{p,n,m}$  can be written in terms of $a_{n,m}$ and $b_{n,m}$ as well.
The last remark is that, the last terms in the expansion of $L_1'$
do follow  the same structures of \reef{uuyy}. Therefore we can conclude that  $f_{p,n,m}$ just in  the expansion of $L_1'$ must be replaced by
\beqa
f_{p,n,m}\rightarrow f_{p,n,m}-f'_{p,n,m}\nonumber\eeqa

\section{Contact terms for $p+2=n$ case}

Let us come to the last part of contact terms. Here all world volume spaces have been covered and apparently there should not be any coupling between gauge field and scalars and Ramond-Ramond. Below one might wonder how we could find a non-zero
coupling between one closed string Ramond-Ramond $(p+1)$-form  and a gauge field and two scalar fields in the world volume of BPS branes.
In order to indeed explore new couplings, first we extract all traces and write down the general form of the non-zero amplitude(which is $A_2$), then we comment on new couplings which can be discovered just by direct super string scattering amplitudes in IIB and IIA theories.
\beqa
{\cal A}_{2}&=&L_2\pi^{1/2}\mu_{p}\frac{16}{(p+1)!}\xi_{2i}\xi_{3j}\bigg\{\epsilon^ {a_{0}\cdots a_{p-1}c}H^{ij}_{a_{0}\cdots a_{p-1}}
(p+1)\bigg(-2us\xi_{1}.k_{2}k_{3c} +ust\xi_{1c}
\nonumber\\&&-2ut k_{3}.\xi_{1}k_{2c} \bigg)
+\epsilon^ {a_{0}\cdots a_{p}}\bigg(2us p^j\xi_1.k_2 H^{i}_{a_{0}\cdots a_{p}}
-2ut p^i\xi_1.k_3H^{j}_{a_{0}\cdots a_{p}}\bigg) \bigg\}
\label{ctermpn233}\eeqa

It is extremely important to notice that, by applying momentum conservation all the terms in the first part of \reef{ctermpn233} do satisfy the only related Ward identity for the gauge field $(\xi_{1c}\rightarrow k_{1c})$. However in order to see the fact that the same result holds for the other terms , we should apply the Bianchi identity as the following :
\beqa
\epsilon^ {a_{0}\cdots a_{p}} \bigg(-p_{a_{p}}(p+1)H^{ij}_{a_{0}\cdots a_{p-1}}-p^j H^{i}_{a_{0}\cdots a_{p}} +p^i H^{j}_{a_{0}\cdots a_{p}}\bigg) &=&dH^{p+2}=0
\nonumber\eeqa

Regarding  above results and in order to have gauge invariance  at leading order of $\alpha'$, we have to find out non-zero couplings for each term in the first part of  \reef{ctermpn233} and then add them up. This rule should be kept for the last two terms in \reef{ctermpn233} as well.
One may expand $ust L_2'$ as below
\beqa
tsu L_2' &=&\pi^{3/2}\bigg(\sum_{n=-1}^{\infty}b_n (s+t)^{n+1}
            +\sum_{p,n,m=0}^{\infty}e_{p,n,m}u^{p+1}(ts)^{n}(t+s)^m\bigg)
 \label {ess}
\eeqa
However, this is not the correct expansion here. Regarding the fact that all Gamma functions are symmetric under interchange of $(s,t,u)$, the final expansion must hold that symmetry as well so the modified expansion has to be taken as
\beqa
tsu L_2' &=&\frac{\pi^{3/2}}{3}\bigg\{\bigg[\sum_{n=-1}^{\infty}b_n (s+t)^{n+1}
            +\sum_{p,n,m=0}^{\infty}e_{p,n,m}u^{p+1}(ts)^{n}(t+s)^m\bigg]\nonumber\\&&+t\leftrightarrow u+s\leftrightarrow u\bigg\}
 \label {esser}
\eeqa
Remember $e_{0,0,0}=0, b_{-1}=1$. First let us try to produce at leading order of $\alpha'$ the second term of the string amplitude in \reef{ctermpn233}, that is
\beqa
\pi^2\mu_{p}\frac{16}{p!}\xi_{2i}\xi_{3j} \epsilon^ {a_{0}\cdots a_{p-1}c}H^{ij}_{a_{0}\cdots a_{p-1}}
 \xi_{1c}
\label{bbb}\eeqa
Note that  if we expand $ustL_2'$ at low energy limit the first term is $\pi^{3/2}$.

One should argue  that the  coupling $(H^{p+2}A\phi\phi)$ has to be derived by field theory manipulation, basically consider  the fact that both  scalar fields could come from either Myers' terms as followed from (58)
or both of them can be resulted in two covariant derivatives in the
 Pull-back  like
 \beqa
\frac{(2\pi\alpha')^2\mu_p}{2(p-1)!}\int d^{p+1}\sigma
\eps^{a_0\cdots a_{p}}
\Tr\left(D_{a_0}\Phi^i\,D_{a_1}\Phi^j\right)
C_{ija_2\cdots a_{p}}
\label{pullback12}
\eeqa

The other possibility is that one scalar can come from pull-back and the second one could come from Taylor expansion as follows

\beqa
 \frac{(2\pi\alpha')^2\mu_p}{p!}\int d^{p+1}\sigma
\eps^{a_0\cdots a_{p}}
\Tr\left(\Phi^j D_{a_0}\Phi^i\right)
\prt_jC_{ia_1\cdots a_{p}}
\label{pullbaxk13}
\eeqa
  As we can clearly see the presence of scalars (which are non-Abelian) in Taylor expansion \cite {Douglas:1997sm,Douglas:1997ch} and pull-back  should be inevitable as was mentioned in \cite{Dorn:1996xk,Hull:1997jc} and  after all we have to extract covariant derivative of scalar to actually receive gauge, scalar and RR couplings. Needless to say we took integrations by parts to indeed combine \reef{pullbaxk13} and \reef{pullback12}, such that their combination is

\beqa
{(2\pi\alpha')^2\over2 p!}\mu_p\int d^{p+1}\sigma
\eps^{a_0\cdots a_{p}}\Tr\left(D_{a_0}\Phi^j\Phi^i\right)
H_{ija_1\cdots a_{p}}
\nonumber\eeqa

If we would open up the covariant derivative of scalar field , we would get two terms but the term involving partial derivative should be dropped as we need to get
 the non-zero coupling of one gauge, two scalars and one RR-$(p+1)$ form so that the ultimate coupling is achieved by

\beqa
 {i(2\pi\alpha')^2\over p!}\mu_p\int d^{p+1}\sigma
\eps^{a_0\cdots a_{p}}\Tr\left(A_{a_0}\Phi^j\Phi^i\right)
H_{ija_1\cdots a_{p}}
\label{esi2378}
\eeqa

It is indeed  a very easy task to observe that \reef{bbb} is precisely reproduced by \reef{esi2378}. Again we want to highlight the point that even we are dealing with $n=p+2$ case and all world volume spaces have been  covered, however, there is a non-zero coupling between RR-$(p+1)$ form and one gauge field and two scalars field but we can no longer write that coupling in terms of field strength of the gauge field.

 \vskip 0.1in

The higher derivative corrections of  \reef{esi2378} can be discovered as

\beqa
&& \frac{i(2\pi\alpha')^2}{3p!}\mu_p \eps^{a_0\cdots a_{p}}H_{ija_1\cdots a_{p}}\bigg(\int d^{p+1}\sigma \bigg[\sum_{n=-1}^{\infty}b_n (\alpha')^{n+1}
\Tr\bigg(\partial_{m_{0}}\cdots \partial_{m_n}A_{a_0} D^{m_{0}}\cdots D^{m_{n}}[\Phi^j\Phi^i]\bigg)\nonumber\\&&+\sum_{p,n,m=0}^{\infty}e_{p,n,m} (\frac{\alpha'}{2})^{p+1}(\alpha')^{2n+m}
\Tr\bigg(\partial_{m_{1}}\cdots \partial_{m_{m}}\partial_{n_{1}}\cdots \partial_{n_{2n}}A_{a_0} \partial^{m_{1}}\cdots \partial^{m_{m}}\nonumber\\&&\times (D_cD^c)^{p+1}[D^{n_{1}}\cdots D^{n_{n}}\Phi^jD^{n_{n+1}}\cdots D^{n_{2n}}\Phi^i]\bigg)\bigg]\bigg)
\label{wwwe}
\eeqa

This prescription can be easily applied to actually get the terms by interchanging $t\leftrightarrow u,s\leftrightarrow u$ and finally we have to add them to \reef{wwwe} as well. Now let us consider the first and third terms in \reef{ctermpn233} and just keep the related leading contact interactions, namely one must employ $usL_2',utL_2'$ expansions in (21) and (22) and keep in mind that the first non zero coefficients are $e_{1,0,0}, e_{0,0,1}$ such that the following terms are leading terms in string amplitude
\beqa
 -\pi^{4}\mu_{p}\frac{16}{6p!}\xi_{2i}\xi_{3j} \epsilon^ {a_{0}\cdots a_{p-1}c}H^{ij}_{a_{0}\cdots a_{p-1}}
\bigg(-2\xi_{1}.k_{2}k_{3c} (t+2s+2u)-2 k_{3}.\xi_{1}k_{2c}(s+2t+2u)\bigg)
\label{rren}\eeqa

\vskip 0.1in

In order to produce all terms in \reef{rren}, one has to work in detail and write down some new  couplings, in the sense that they do not come from pull-back, Taylor or Myers' terms. Therefore the first term above can be reproduced by the following coupling:

\beqa
&& -{4\pi^2(2\pi\alpha')^2\over 6p!}\mu_p\int d^{p+1}\sigma
\bigg(\partial_bA_aD^bD^a\Phi^iD_c\Phi^j+2\partial_bA_a D^a\Phi^iD_c\partial^b\Phi^j\nonumber\\&&+2 A_a D^a\partial_b\Phi^iD_c\partial^b\Phi^j-2\leftrightarrow 3\bigg)
H^{(p+2)}_{ija_0\cdots a_{p-1}}\eps^{a_0\cdots a_{p-1}c}
\label{newwq}
\eeqa

Now we get to the point which has been emphasized, namely in order to produce all the first three terms in \reef{ctermpn233}, one has to add \reef{esi2378} and \reef{newwq} together, and generalizing to all orders can be easily done with making use of (71).

\vskip 0.1in

Now let us consider the last two terms in \reef{ctermpn233} and just keep the related leading contact interactions. Having set the  $usL_2',utL_2'$ expansions from (21) and (22) we get
\beqa
 -\pi^{4}\mu_{p}\frac{16}{6(p+1)!}\xi_{2i}\xi_{3j} \epsilon^ {a_{0}\cdots a_{p}}
\bigg(2 p^j\xi_{1}.k_{2} H^{i}_{a_{0}\cdots a_{p}} (t+2s+2u)-2 p^i k_{3}.\xi_{1} H^{j}_{a_{0}\cdots a_{p}}(s+2t+2u)\bigg)
\label{mmnb}\eeqa

\vskip .1in

By applying the correct higher derivative corrections , it is easy to show that
some new couplings must be taken in to account, such that the first term above can be reproduced by the following coupling:

\beqa
&& {4\pi^2(2\pi\alpha')^2\over 6(p+1)!}\mu_p\int d^{p+1}\sigma
\bigg(\partial_b A_a D^b D^a\Phi^i \Phi^j+2 \partial_b A_a D^a\Phi^i D^b\Phi^j
\nonumber\\&&+ 2A_a D^bD^a\Phi^iD_b\Phi^j-2\leftrightarrow 3\bigg)
\partial_jH^{i}_{a_0\cdots a_{p}}\eps^{a_0\cdots a_{p}}
\label{newwqvv}
\eeqa

Again we want to highlight the point that these new couplings do not come from pull-back or Taylor or Myers' terms. Having used (71), we can easily generalize above couplings to actually get all contact interactions to all orders of $\alpha'$.

\vskip .2in

However, it turns out that it is better to write down the closed form of contact terms  to all orders
of $\alpha'$ rather than producing them order by order in $\alpha'$. Thus consider the following terms in string amplitude

\beqa
&& -\pi^{2}\mu_{p}\frac{16}{p!}\xi_{2i}\xi_{3j} \epsilon^ {a_{0}\cdots a_{p-1}c}H^{ij}_{a_{0}\cdots a_{p-1}}
\sum_{p,n,m=0}^{\infty}e_{p,n,m}\bigg(-2\xi_{1}.k_{2}k_{3c} t^p (su)^n (s+u)^m
\nonumber\\&&-2\leftrightarrow 3\bigg)
\label{rre}\eeqa

Their closed form  can be precisely obtained to all orders of $\alpha'$ by the following coupling:

\beqa
&& \frac{\mu_p(2\pi\alpha')^2}{p!} \sum_{p,n,m=0}^{\infty}e_{p,n,m}(\frac{\alpha'}{2})^{p}(\alpha')^{2n+m+1}\int d^{p+1}\sigma
\bigg((D_a D^a)^p D_{a_{1}}\cdots D_{a_{m}}[\partial_{a_{1}}\cdots \partial_{a_{n}} A_a \nonumber\\&&\times  D_{a_{n+1}}\cdots D_{a_{2n}}D^a\Phi^i]D_{a_{p}}
D^{a_{1}}\cdots D^{a_{m}}
D^{a_{1}}\cdots D^{a_{2n}}\Phi^j+2\leftrightarrow 3\bigg)
H^{ij}_{a_{0}\cdots a_{p-1}}\eps^{a_{0}\cdots a_{p}}
\label{newwnmb}
\eeqa

Finally the rest of the terms in string amplitude to all orders are verified as
\beqa
 -\pi^{2}\mu_{p}\frac{16}{(p+1)!}\xi_{2i}\xi_{3j} \epsilon^ {a_{0}\cdots a_{p}}\sum_{p,n,m=0}^{\infty}e_{p,n,m}
 t^p (su)^n (s+u)^m\bigg(2 p^j\xi_{1}.k_{2} H^{i}_{a_{0}\cdots a_{p}}-2\leftrightarrow 3 \bigg)
\label{mmnb}\eeqa

Eventually  by applying the same methodology as discussed in the body of the paper, one can easily show that \reef{mmnb} would be produced by the following couplings:

\beqa
&& \frac{\mu_p(2\pi\alpha')^2}{(p+1)!} \sum_{p,n,m=0}^{\infty}e_{p,n,m}(\frac{\alpha'}{2})^{p}(\alpha')^{2n+m+1}\int d^{p+1}\sigma
\bigg((D_a D^a)^p D_{a_{1}}\cdots D_{a_{m}}[\partial_{a_{1}}\cdots \partial_{a_{n}} A_a \nonumber\\&&\times  D_{a_{n+1}}\cdots D_{a_{2n}}D^a\Phi^i]
D^{a_{1}}\cdots D^{a_{m}}
D^{a_{1}}\cdots D^{a_{2n}}\Phi^j-2\leftrightarrow 3\bigg)
\partial_j H_{i a_{0}\cdots a_{p}}\eps^{a_{0}\cdots a_{p}}
\label{newwnmb}
\eeqa

\vskip .1in

\section{Conclusion}

First of all by applying   conformal field theory methods we discovered the complete result  of the amplitude of one Ramond-Ramond, one gauge field and two scalar  fields for all kinds of $p$ and $n$ in II string theory. The motivation for carrying out this long computation was that, we must have the complete form of the amplitude to be able to proceed to explore the closed form of  new Wess-Zumino couplings to all orders in $\alpha'$ for various cases. We have also performed all SYM vertex operators to all orders in $\alpha'$. Remember that due to closed string Ramond-Ramond the general form of these new couplings  with their exact coefficients should be confirmed just by direct S-Matrix computations not any other tool like T-duality transformation to the previous calculations. The results of this paper can not be found for example from $ <V_C V_A V_A V_{\phi}>$ because of the fact that $C$-vertex operator does not have Winding modes in its form, which means that all the terms including $p^i, p^j$ of this paper have net been showed up in $ <V_C V_A V_A V_{\phi}>$. We have shown that the amplitude of $CA\phi\phi$  has infinite massless poles in various channels.  Namely making use of the all order two gauge two scalar couplings \cite{Hatefi:2012ve}, we were able to match all infinite massless gauge poles in $(t+s+u)$-channel in string theory amplitude with field theory computations. We have also produced all infinite massless scalar poles in t,s-channels in both field and string theory sides. Apart from those things, we showed that the amplitude has again infinite  massless gauge poles in u-channel for $p=n+2$ case and then we went through new WZ couplings for this case like $\Tr(C_{p-3}\wedge F\wedge D\phi^i\wedge D\phi_i)$ and all its infinite higher derivative corrections have been explored in (37).

\vskip .1in

It has been eventually clarified that the couplings of two scalars and one gauge field can not hold any corrections, thus all non-leading (gauge/scalar) poles have provided the most needed information to indeed get the all order $\alpha'$ higher derivative corrections to $\Tr(C_{p-3}\wedge F\wedge F)$ and $\bigg(\Tr(\phi^j\phi^i)\partial_j H_{ia_{0}\cdots a_{p}}+(p+1)\Tr(\partial_{a_{0}}\phi^j\phi^i)H_{ija_{1}\cdots a_{p}}\bigg)$.

\vskip 0.1in

 We also found further results, basically in order to produce all infinite $t,s$-channel scalar poles we found (64). To get all infinite contact interactions
for $p=n$ case , (67),(71) are derived. New couplings and all their infinite $\alpha'$ corrections are discovered in (74)-(83).
In order to actually derive  all contact terms for $p+2=n$ case, we have also obtained several new Wess-Zumino couplings in (93),(95) with all their infinite corrections in (97) and (99) and also (91) is derived to all orders  . These new interactions which are neither inside Myers'terms nor within pull-back/Taylor expansion  must be looked for only by performing  direct string computations. It would be nice to perform either $CAA\phi\phi$ or $CAAA\phi\phi$ to get some more information , remove some of the ambiguities which are addressed in this paper and finally to see whether or not covariant derivatives should be kept inside the new couplings of this paper.


\section*{Acknowledgments}
It is my pleasure to thank Rob Myers for several useful discussions and for sharing his ideas on some new WZ couplings with me. The author would like to thank  J.Polchinski, K.S.Narain ,E.Gava and F.Quevedo for valuable discussions. He acknowledges  E.Witten, G.W.Moore, W.Taylor  and J.Maldacena  for useful discussions during Strings 2012 held at Munich also thanks L. \'Alvarez-Gaum\'e, N.Lambert and G.Veneziano for several valuable comments. He also thanks Perimeter Institute for warm hospitality during his visit where some part of this work was taken place there .


\end{document}